\documentclass[letterpaper,11pt]{article}

\usepackage[numbers]{natbib}
\usepackage{lineno}
\usepackage{graphicx}
\usepackage{verbatim}
\usepackage{xspace}
\usepackage{amsmath}
\usepackage{setspace}
\usepackage{enumitem}
\usepackage{epigraph}
\begin{document}

\doublespacing

\thispagestyle{empty}

\vspace{30mm}
\begin{center}
{\LARGE \textbf{Simplification, innateness, and the absorption of meaning from context: how novelty arises from gradual network evolution}}
\\
\vspace{3mm}
{\Large Adi Livnat\\
\bigskip
Department of Evolutionary and Environmental Biology and Institute of Evolution, University of Haifa, Mount Carmel, Haifa 3498838, Israel. 

\vspace{5mm}
alivnat@univ.haifa.ac.il}
\end{center}

\newpage

\section*{Abstract}

The theory of interaction-based evolution argues that, at the most basic level of analysis, there is a third alternative for how adaptive evolution works besides \textit{a}) accidental mutation and natural selection and \textit{b}) Lamarckism, namely, \textit{c}) information provided by natural selection on the fit between the organism and its environment is absorbed by non-accidental mutation. This non-accidental mutation is non-Lamarckian yet useful for evolution, and is due to evolved and continually evolving mutational mechanisms operating in the germ cells. However, this theory has left a fundamental problem open: If mutational mechanisms are not Lamarckian---if they are not ``aware'' of the environment and the macroscale phenotype---then how could heritable novelty be due to anything other than accidental mutation? This paper aims to address this question by arguing the following. Mutational mechanisms can be broadly construed as enacting local simplification operations on the DNA in germ cells, along with gene duplication. The joint action of these mutational operations and natural selection provides simplification under performance pressure. This joint action creates from preexisting biological interactions new elements that have the inherent capacity to come together into unexpected useful interactions with other such elements, thus explaining nature's tendency for cooption. Novelty thus arises not from a local genetic accident but from gradual network-level evolution. Many empirical observations are explained from this perspective, from cooption and gene fusion at the molecular level, to the evolution of behavior and instinct at the organismal level. Finally, the nature of mutational mechanisms and the need to study them in detail are described, and a connection is drawn between evolution and learning. 

\bigskip
\noindent \textbf{Keywords:} 
Evolvability, learning, instinct, stereotypy, genetic assimilation, evolution of language, parsimony.  

 \newpage

\epigraph{\itshape The problem is not to choose the correct scale of description, but rather to recognize that change is taking place on many scales at the same time, and that it is the interaction among phenomena on different scales that must occupy our attention.}{---Simon A. Levin, 1992.}

\section{Introduction}

\begin{figure}[pdf] \centering{
\includegraphics{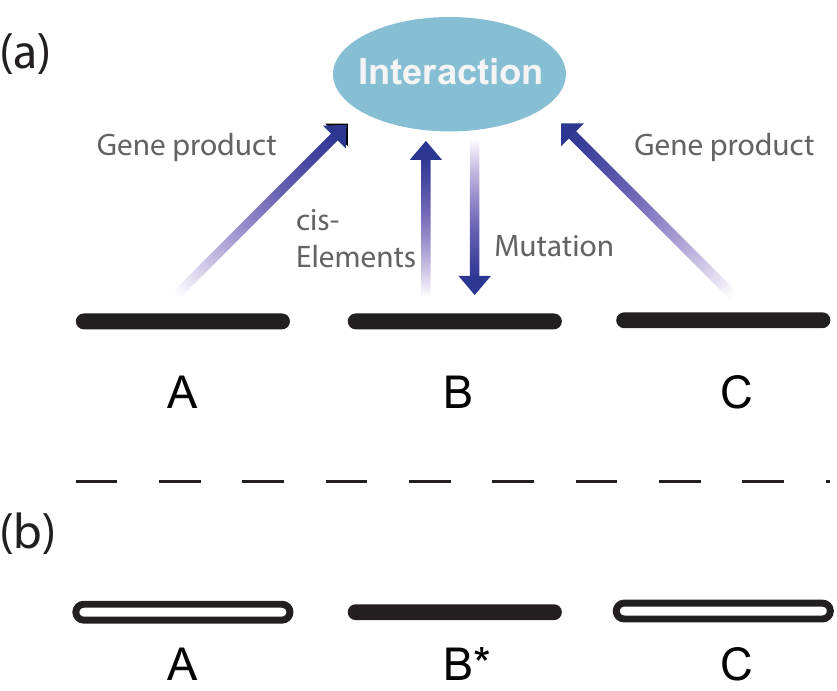}
}
\caption{Mutation as a biological process, from \cite{Livnat2013}. a) In this figure we see three loci coming together in a biological interaction through gene products and cis elements. This part of the figure merely represents schematically the gene regulation and interaction that are key to our understanding of molecular and cellular biology. What is new about the figure is that we have not yet fully considered the possibility that there could be a mutation arrow too, i.e., that mutation is an outcome of genetic interactions in a heritable mode; i.e., that much like genes interact in influencing a classical trait, like the eye or the ear, they also interact in influencing genetic change. Note that the figure purposely leaves open the particulars of the biochemical mechanisms involved, as there may be many such mechanisms, and that ``mutation'' is broadly construed to mean any heritable genetic change. These may involve not only DNA changes but also epigenetic changes. b) Mutation as an event of information flow and computation changes many things in our conceptualization of evolution. Particularly, the biological process of mutation creates from the combination of interacting alleles across loci a new heritable piece of information---a new mutation---a new allele, B*. Even though the particular combination of interacting alleles will sooner or later disappear due to the sexual shuffling of the genes, information from it can be transmitted to future generations through the mutation. In this manner, the problems of the role of sexual recombination and of the nature of mutation may be tied together.}  
\label{muwriting}
\end{figure}

The theory of interaction-based evolution \cite{Livnat2013} argues that the mutations that drive adaptive evolution under selection are not local accidents occurring to the genome. Instead, they result from the action of evolved and continually evolving complex biological mechanisms  \cite{Livnat2013}  
and are therefore affected by genetic interactions across loci. It follows that mutation combines information from alleles across loci and writes the result of the combination into one locus---the locus of the mutation \cite{Livnat2013}. The schematic figure that describes this nature of mutation (Figure \ref{muwriting}a) is much like that which would represent gene interaction and regulation, except that the outcome of the action in this case is genetic change. ``Mutation'' here is broadly construed to encompass not only DNA mutations but also epigenetic changes. 

\begin{figure}[pdf] \centering{
\includegraphics{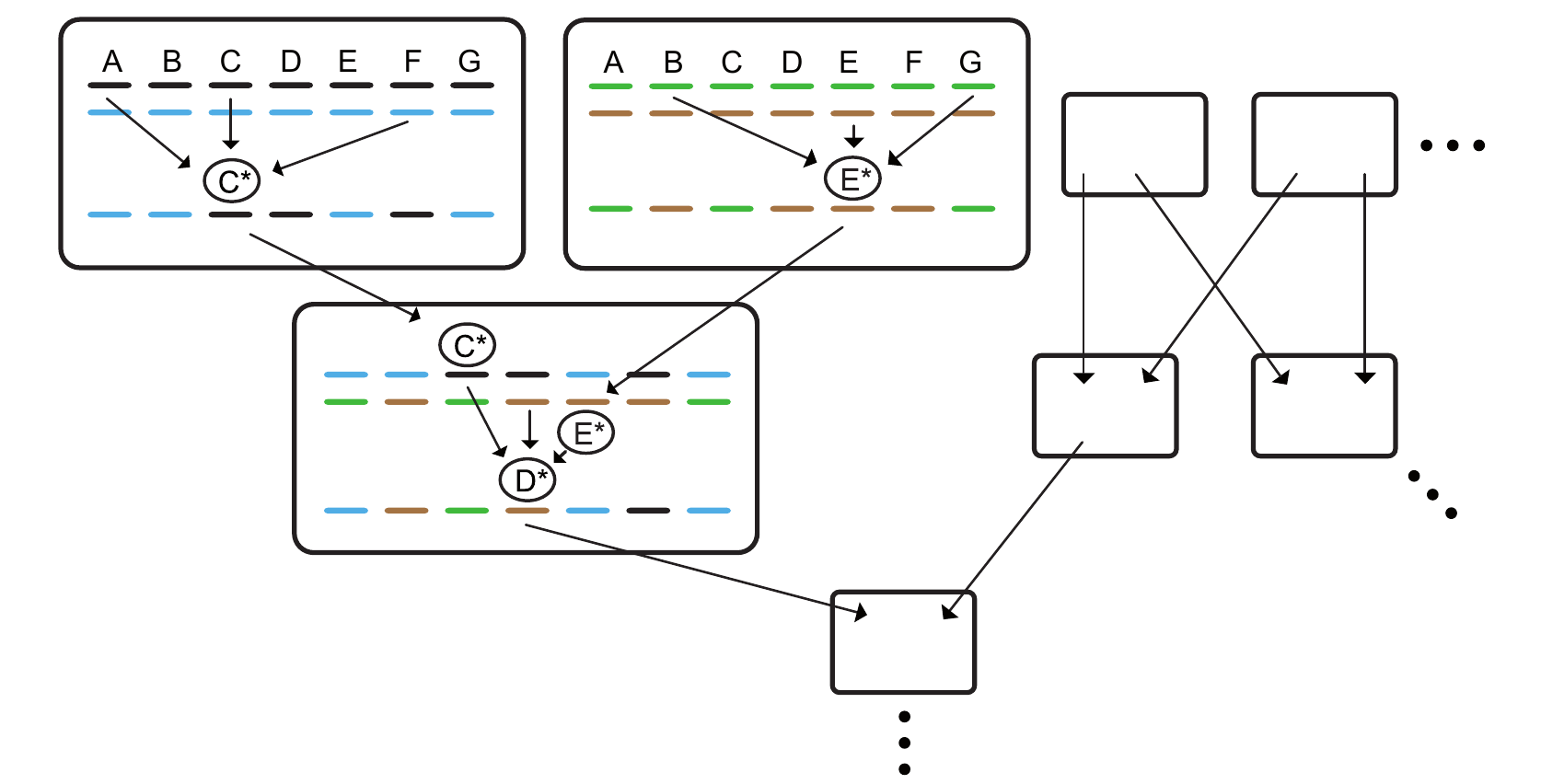}
}
\caption{Mutation as an event of information transmission and computation creates a network of information flow through the generations, from \cite{Livnat2013}. Each box represents an individual, and in each box, the two sets of lines at the top represent that individual's diploid genotype (genes A through G), and the set of lines at the bottom represents a haploid genotype transmitted through the gamete. For the sake of demonstration, a small number of mutational events due to interactions between genes is shown in two parents and an offspring (large boxes), although many mutations occur in other genes and in other individuals at the same time. For example, C* represents a mutation in one of the alleles of gene C. Because the \textit{output} of a mutational event in one generation---namely the mutation itself (e.g., C*)---can serve as an \textit{input} into mutational events at later generations (e.g., the event creating D*), non-accidental mutation creates a network of information flow and computation over the generations, from many genes into one and from one gene into many, as well as from many individuals into one and from one individual to many.} 
\label{flowovergenerations}
\end{figure}

Moving to the population level, we see that the outcome of a mutational event in one generation---namely the mutation itself---can serve as an input into mutational events at later generations \cite{Livnat2013}. Therefore, mutations create a network of information flow across the genome and through the generations (Figure \ref{flowovergenerations}) \cite{Livnat2013}. This suggests at the outset a process by which the genome can evolve as a cohesive whole \cite{Mayr1963,Livnat2013}.

This view immediately affects how we conceptualize fundamental questions in evolution, such as the question of the role of sex in evolution \cite{FeldmanEtal1997}. A 
layman's intuition has been that, since natural selection acts on individual variation, the vast number of different genetic combinations generated by sex facilitates adaptive evolution. However, this answer has been incomplete from a theoretical perspective because, just as sex puts together these combinations, it also breaks them down: they are not heritable. However, if mutation is not simply a local accident, but instead encapsulates a flow of information across loci, then although individual genotypes are transient, they can have effects on future generations through the mutations that are derived from them (Figure \ref{muwriting}b) \cite{Livnat2013,LivnatPapadimitriou2016}, and the original intuition holds in some sense. Such information flow through mutation enables a situation where selection evaluates each individual \textit{as a complex whole}, and information from that individual as a complex whole is passed on by mutations precisely in accord with the individual's fitness \cite{Livnat2013}\footnote{We are no longer restricted to the effective transmission of additive genetic effects.}. 

Another such question is the nature of mutation. Recently, evidence has been accumulating showing that mutational events are complex and involve genetic information and biological mechanisms \cite{Livnat2013}. From a traditional standpoint, these complex influences on mutation are seen as happenstantial and do not command attention. In contrast, interaction-based evolution argues that they are at the heart of the evolutionary process.

By putting together the problem of the role of sex in evolution, the question of the nature of mutation and more, interaction-based evolution has put together many questions and observations previously disconnected and has raised multiple predictions and directions for future research \cite{Livnat2013}. However, it has left a fundamental problem open. The traditional view takes random mutation to be the ultimate source of heritable innovation and creativity in evolution: random mutation invents, and natural selection selects\footnote{Even the evolvability approach \cite{Koonin2011,WagnerAltenberg1996}, which allows for the evolution of mechanisms affecting mutation \cite{Kimura1967,Leigh1970,AltenbergFeldman1987}, still relies either implicitly or explicitly on accidental mutation at the origin of things, and assumes that evolvability mechanisms are merely later add-ons to the core process of random mutation and natural selection, ones that play a facilitatory but not a fundamentally necessary role.}. 
However, if mutation is not accidental and never was, then what is the ultimate source of heritable novelty? 

In particular, interaction-based evolution does not admit Lamarckism---it does not admit a mechanism that senses a phenotypic need in multicellulars through interaction with the environment and translates that need into the required genetic change. But if the influences on mutation are not ``aware'' of the environment and the phenotypic need, then how could the ultimate source of heritable novelty in evolution be anything other than random mutation? This paper will propose an answer, thus completing the replacement to random mutation at a conceptual level that started with the first interaction-based evolution paper \cite{Livnat2013}.  

Inspiring, long-term efforts by Wagner and colleagues have shown that network-level evolution is key to innovation (e.g., \cite{LynchEtal2011,EmeraWagner2012a,EmeraWagner2012b,Wagner2014}). To answer the question above, I will continue these efforts in the direction of interaction-based evolution. I will propose here the following. 
\textit{i}) Novelty arises from gradual network-level evolution. \textit{ii}) The phenotypic meaning of a genetic element is \textit{gradually absorbed} from the network in the course of network-level evolution and is not bequeathed to it by a local genetic accident. \textit{iii}) Molecular cooption---i.e., the case where a preexisting genetic element comes to be used in a new context---is not simply an outcome of stochastic events but is an outcome of a \textit{gradual} process of network-level evolution, where non-accidental mutations pave the way and predispose the genome to cooption. \textit{iv}) This process of gradual network-level evolution and the
fact that the phenotypic meaning of a mutation comes from context rather than arises anew based on a specific function \textit{per se} also explain the evolution of innateness, previously known as the problem of the ``inheritance of acquired characters.'' In this connection, 
we will see that automatization is at the essence of the evolutionary process. \textit{v}) Simplification and complexity are connected: While selection puts a pressure for organismal level performance, there exists in addition genetic simplification pressure due to mutational and recombinational mechanisms. Together, the pressures for performance and simplification drive the evolution of complexity and novelty, surprisingly connecting simplicity and complexity. In particular, elements simplified under performance pressure are expectedly unexpectedly useful: they have the inherent capacity to come together in interaction with other such elements and thus become useful in unexpected, novel ways. \textit{This inherent ability, which accounts for cooption, is the source of novelty in evolution.} \textit{vi}) Evolution is driven at the molecular level by evolved and continually evolving mutational mechanisms that implement useful operations, much like Hebbian learning and other non-random operations are thought to be useful in learning. A search for these mutational mechanisms, both empirical and theoretical, needs to begin. 

The paper is organized as follows. The next section will describe the nature of network-level evolution. Section \ref{simplification} will introduce the idea of simplification under performance pressure. Together, these two sections will propose how non-accidental mutations could be useful for evolution yet be non-Lamarckian, and how novelty arises. Section \ref{innateness} will then bring a large number of empirical observations in support of the view proposed here. These will be observations on the evolution of behavior at the organismal scale. Of particular importance will be subsection \ref{backwardwalking}, where all the concepts developed will come together in an empirical example with an emphasis on the evolution of novelty. Finally, section \ref{newview} will revisit the molecular level in light of the concepts developed, discuss the nature of mutational mechanisms and draw a connection between evolution and learning, including machine learning, thus underscoring the importance of the algorithmic lens \cite{Papadimitriou2007,Karp2011} for our understanding of evolution. 
 
\section{A contextual view of genetics}
\label{contextual}

Due to the molecular biological revolution, it has become clear that the same or similar genetic element can be seen in two or more different genetic contexts within the same species or in different species \cite{KirschnerGerhart2006}. This means that, over evolutionary time, a molecule can change the context in which it serves---it can be ``coopted.'' For example, the frog toxin caerulein has been independently coopted from the homologous gastrointestinal peptide hormones cholecystokinin and gastrin, with whose action it interferes in the affected animals \cite{RoelantsEtal2010,BowieEtal2006}. And proteins involved in cellular stress response, like the small heat shock proteins \cite{IngoliaCraig1982}, have often been coopted as light refracting crystallins in the lens of the eye, an avascular tissue presenting harsh biophysical conditions \cite{TrueCarroll2002}. Indeed, ``Cooption,'' ``opportunism,'' or ``tinkering'' \cite{Jacob1977,GraurLi2000} is so important that it has been called ``the paradigm of molecular evolution'' \cite{GraurLi2000}. But how does cooption happen? Does a genetic sequence just jump one day by accident from one locus to another and acquires a new use? 

Traditional discussions admit but do not explain shifts in the context of usage of a genetic element or a phenotype. 
In them, natural selection is limited to building up one independent or additive contribution to fitness on top of another toward advancement in the same adaptation. This provides no explanation for cases where an element is first used in one context and then in another, beyond saying that they are due to chance. This paper will begin to fill this gap, by delving into the question of what makes it so that evolution is capable of producing building blocks that, combined with other elements in a network, produce novel functionality. 

As will become relevant soon, we often see that fusion accompanies cooption. For example, members of the cyclophilin family, which have been found in bacteria, fungi, plants and animals \cite{TrandinhEtal1992}, have a peptidyl-prolyl cis-trans isomerase activity which allows them to participate in diverse biological processes in all subcellular compartments, from protein translocation across membranes, to mitochondrial function, to control of transcription, and more (see \cite{ColganEtal2000} and references therein); and it is the presence of different additional domains in the different family members that specifies their unique localizations and interactants \cite{ColganEtal2000}. The next section will examine a particular fusion involving a cyclophilin family member, cyclophilin A. 

\subsection{Cooption at the molecular level is due to a gradual process}
\label{trimcypfusion}

We will now see two motivating examples, one from molecular evolution and one from phenotypic evolution. 

Cyclophilin A (CypA) is a highly abundant cytosolic protein \cite{HaendlerHofer1990} that, among its various activities, potently binds several retroviral capsids, including HIV-1 \cite{KaessmannEtal2009}. TRIM5 is a restriction factor that recognizes and inactivates incoming retroviral capsids \cite{VirgenEtal2008}. A copy of the \textit{CypA} gene has retroposed into the \textit{TRIM5} gene independently in 
at least two different simian lineages \cite{VirgenEtal2008,NisoleEtal2004,SayahEtal2004,LiaoEtal2007,BrennanEtal2008,WilsonEtal2008,NewmanEtal2008}, and the resulting TRIM5-CypA fusion protein appears to provide strong protection against certain lentiviruses   \cite{NisoleEtal2004,SayahEtal2004}. The curious nature of these independent fusions has been noted \cite{VirgenEtal2008,Livnat2013}: not only is a repeated fusion event even more surprising from a traditional perspective than a repeated point mutation (there are many more possibilities of fusion, making repeated fusion by chance even less likely), there are many other \textit{TRIM} genes, and tests of artificial fusions of the CypA domain to some TRIM motifs have shown that they too can provide retrovirus protection \cite{ZhangHatziioannouEtal2006,YapEtal2006,YapEtal2007}, yet \textit{TRIM5} specifically repeats in both fusions mentioned above \cite{VirgenEtal2008}. Furthermore, it has been suggested that genetic factors have influenced the probability of the fusion, such as the extensive transcription of \textit{CypA} in the germline \cite{KaessmannEtal2009,JohnsonSawyer2009,VirgenEtal2008,Livnat2013}.

The current theory argues that this fusion (and others like it) was not due to a sudden, chance event, but rather was the culmination of a gradual genetic and phenotypic evolutionary process that led to it. Minor genetic changes have accumulated, predisposing the genome to the appearance of the fusion, and thus accounting for the fact that it appeared independently multiple times. Furthermore, I argue that \textit{TRIM5} and \textit{CypA} interacted with each other prior to their fusion. Thus, the fusion did not cause \textit{TRIM5} and \textit{CypA} to interact to begin with, but rather was led by their preexisting interaction. 

Furthermore, I hypothesize a specific mechanism that promotes such fusions. Two genes that work together in the soma in a particular context likely are transcribed at the same time. Because they may share cis elements and transcription factors that activate them, information indicating that they work together in the soma is likely present in the DNA and  accessible in the germline, in particular to the  transcriptional machinery. The two genes may be transcribed in the germline at the same time, making it so that the chromatin will be open at both loci at the same time. And since reverse transcription occurs in the germline \cite{BrosiusTiedge1996}, it will be more likely to land a copy of one of these genes next to the other in the DNA. Other steps may further facilitate the fusion, such as trans-splicing prior to reverse transcription. Interestingly, the fact that transcription is promiscuous in the germline allows any genes---somatic as well as germline genes---to participate in this mechanism \cite{Livnat2013}. 

One may think that it just so happens that the genetic system allows for such mechanisms, or that they are fortuitous ``accidents.'' However, following \cite{Livnat2013}, I argue that, rather than being happenstantial, mechanisms of this sort are of much significance. In particular, the mechanism abovementioned is reminiscent in a certain respect of Hebbian learning in neuroscience (Stephen Pacala, personal communications)\footnote{The connection between evolution and learning will be further elaborated on in section \ref{evolutionaslearning}}. According to Hebbian learning \cite{Hebb1949}, when one neuron persistently participates in causing another to fire, the strength of the connection between them is increased, making it so that neurons ``wire together if they fire together'' \cite{LowelSinger1992}. Similarly, here, 
I argue that \textit{copies of genes that are used together are fused together}. Or, to be more precise, copies of genes that are persistently used together in a new context are more likely to be fused. 
Note that this Hebbian-learning--like genetic operation is implemented by the mutational mechanism itself. 
This contrasts with a recent proposal involving Hebbian learning in evolution without invoking non-accidental mutation \cite{WatsonSzathmary2015} and accords with the principle of interaction-based evolution, according to which the mutations relevant for adaptive evolution are non-accidental.  

\subsection{Cooption at the phenotypic level is due to a gradual process} 
\label{incitingceremony}

Examples of cooption and fusion are also apparent at the phenotypic level. Consider the inciting ceremony in ducks \cite{Lorenz1971,Lorenz1958,Lorenz1966}. In the European common shelduck (\textit{Tadorna tadorna}), when the female is standing near her mate, her aggression instinct is triggered by the presence of neighbors, and she may run toward them with her neck stretched, which is the threat posture in ducks 
\cite{Lorenz1966}. As she approaches them she naturally becomes fearful,  turns around and flees back toward her drake\footnote{This to-and-fro movement is not surprising, as it is very common in territorial disputes across species of birds, fish and mammals.}. Approaching her drake, the former instinct is triggered again. In those cases where her breast is still facing him, she turns her neck back to threaten the neighbors over her shoulder. This behavior by the female can incite her mate to attack the neighbors. Note that the angle between the neck and the body of the female is entirely dependent here on the situation---her body orientation is due to the location of the drake and her neck orientation is due to the location of the neighbors \cite{Lorenz1966}. 

Lorenz followed the homologues of this behavior in other duck species and suggested that the to-and-fro movement such as seen in the common shelduck has gradually become ritualized, so that, in the ritualized forms, the female does not perform the to-and-fro but stands near her drake; and, most interestingly, the two elements of orienting the body toward the male and stretching the neck over the shoulder toward the neighbors---which in the non-ritualized forms  
are triggered separately by the environment---have become welded together \cite{Lorenz1971,Lorenz1958,Lorenz1966}. For example, in the East European-Asiatic ruddy sheldrake
(\textit{Tadorna ferruginea}), the neck and body orientations are still controlled separately, but in most of the cases the female stands with her breast to the drake and her neck pointing backwards (and very rarely this behavior may be performed without a neighbor present) \cite{Lorenz1966}. And in the mallard (\textit{Anas platyrhynchos}), the same breast-to-the-male-and-pointing-backwards is observed, but now this posture is compulsory and, at high excitation, which turns the instinct on (the same relationship between excitation and activation of instinct exists for many other instincts), the female is compelled to turn her neck over her shoulder even if that means that the neck moves away from the neighbor \cite{Lorenz1958}. Thus, two elements of behavior, previously triggered separately by two separate environmental triggers, have become welded together and triggered as one. Finally, in the golden-eye (\textit{Bucephala}), where the movement is highly ritualized (see below), the presence of a conspecific is not even required \cite{Lorenz1966}. 

Interestingly, along with the evolutionary change of form of the behavior, there has been also an evolutionary change of meaning. In the species with the less-ritualized form, the behavior has the effect of inciting and is related to territorial behavior. However, note that it already has in it an element of pair-bonding, or team work. In the more ritualized cases, this pair-bonding meaning has moved to the fore: in the mallard, though it sometimes still elicits a demonstration of attack by the male, inciting serves mostly as an invitation to pair-bond; and in the golden-eye, the inciting has become almost entirely independent of the presence of neighbors, and takes a highly ritualized, exaggerated and rhythmic form of neck movements over one shoulder and then the other (and rhythmic movement is indicative of highly ritualized behaviors in general). 

It is due to the highly surprising nature of this example and others that Lorenz has been accused of Lamarckian thinking. However, many examples of this sort exist, and we will see that they are explained not by Lamarckism but by network-level evolution (sections \ref{verbalmodel}, \ref{innateness}). What is important to notice in the two examples discussed so far is as follows. In both of them, we see a \textit{gradual process} 
arising from \textit{preexisting interactions}.  
A novel phenotype (the fused protein in one case, the ritualized display in the other) arises from the change in context in which preexisting elements  
(preexisting genes, movements) are embedded. 
In fact, what was once an interaction has now become an object: in the case of \textit{TRIM5-CypA}, a hypothesized interaction between two separate genes is succeeded by a gene fusion; and in the evolution of the inciting ceremony, two separate behavioral responses to two separate environmental triggers (orienting the body toward the drake and threatening the neighbors over the shoulder) has now become fused into a new  
instinct. Thus, the source of novelty is in system-level changes. In both cases, novelty arises \textit{not} from a point-wise change, \textit{not} suddenly and \textit{not} out of thin air.  

Among else, we also see local simplification in both cases:  in simians, what previously required the separate transcription of two genes now requires the transcription of one, and in ducks, a roundabout to-and-fro behavior has now turned into a stationary clear display. These aspects and more will be explored in-depth in this paper, leading to novel insights on the fundamental nature of evolution and to a macroscale-view of the theory of interaction-based evolution \cite{Livnat2013}. 

\subsection{Network evolution and its operators}
\label{verbalmodel}

I will now propose a verbal model that ties shifts in context to network-level evolution. The model is purposely described at a high level because its role is to elucidate concepts, not to provide mechanistic detail. 

Consider that in the course of genetic evolution, the network of genetic interactions gradually changes as a whole. Many changes take place over the genome and over time, and these changes interact. 
This process involves regulatory changes that can rewire the genetic network \cite{Carroll2005}, such as movements of transposable elements carrying with them cryptic enhancer/promoter sites and multiple mutations activating those sites, for example \cite{LynchEtal2011}. 
But even a regulatory change that at first sight appears only to change the strength of an already existing connection between two nodes---e.g., to increase the effect of a regulator on its target---can effectively cause rewiring; because there is no sharp boundary between the case where the regulator has a negligible effect on its target (in which case the two nodes can be said to be effectively disconnected) and the case where it has a non-negligible effect (where the two nodes can be considered to be connected). 

Rewiring means that, in the course of evolution, the connections between some nodes on the network become tighter and the connections between other nodes become weaker, and recognizing it is important. When 
the connections between nodes become tighter, they come to be regulated more and more as one unit, and a new module arises. What in the beginning may be two separate elements regulated by two separate lines of control can gradually come under one line of control. As will be understood later, this change represents the arrival of a new automatic unit. Furthermore, when  
this coming together of genes is preceded by the duplication of those genes and their regulatory elements, this new module does not arise at the expense of previous ones, but represents a total increase in the number of modules; and together with this increase in the number of modules comes an \textit{increase in the extent of higher-level interactions} between modules (since all the modules must ultimately come together into one organism, and now there are more of them\footnote{``Number'' of modules and ``more'' modules could be put in quotations because modules do not have a precise number, as they ultimately grade into each other, indeed because they have to be connected to each other. The definition of a module used in the literature is a fuzzy one and rightly so: it is a set of genes that interact more closely with each other than with other genes, even though to interact with the ``outside,'' at least some of its members have to have just as strong a connection to members outside of the module. However, the fact that we cannot perfectly count the total number of modules is an inherent characteristic of the process: it allows new modules to gradually form.}.) 

While the term ``module'' usually refers to a set of tightly interacting genes, a rather basic module or unit is an exon; and since exons in separate loci may interact through trans-splicing, or through protein-protein interactions, etc., the same kind of process can cause the coming together of two previously interacting exons into a gene, or gene fusion. Such a fusion may be long in the making. This shows us a case where a new elementary unit evolves from an interaction---from a process---and where a process can become an object---a gene. And as an object, it begins to accept the kind of operations that the system can apply to other objects. It is now interacting directly and indirectly with many other units. 

A critical point in the above now calls for reflection. It takes time for two elements to undergo separate regulation and transcription in order to come together later into a functional unit or interaction. But when they come together evolutionarily into one genetic unit, regulated as one and performing through one product, this time is cut to zero. Previously, the joint effect of these two elements came into being as developmental interactions do; now it is ``innate''---it is a gene. It no longer needs to be constructed from more elementary units, and it exerts its effect in interaction with other (now-peer) elementary units, in whose context it has phenotypic meaning. The emphasis here is not on the actual amount of time cut, but on the local simplification of the network. 

Thus, in the gradual fusion of two elements into one, we see a sense of evolutionary acceleration of developmental interactions; and if this fusion is preceded by the copying of those two elements, we see at the same time 
an increase in the ``genetic vocabulary,'' which comes together with an increase in the extent of higher-level interactions---an increase in complexity. 

Having thus formed a clear view of acceleration and the arising of new interactions with the help of the gene fusion case, it is important to step back again and observe these two aspects in the big picture.  
It is enough to consider the copying of modules and the changing of regulatory connections (prior to considering actual gene fusion) in order to notice that these changes of connections can be seen from two angles: 
When we look at the lower levels of organization---at the tightening of connections between nodes---we see an increase in innate abilities. When we look at the higher levels of organization---at the increase in the extent of interactions between modules due to the appearance of new modules---we see an increase in the complexity of the life-form, the phenotype. Importantly, these are two facets of one integrated process: the new parts observed at the lower levels (which are due to constriction) and the new whole (which is due to the increase in the extent of high-level interactions) \textit{coevolve}. The novelty comes from a network-level change, not from a sequence of independent, atomistic changes. And, as will be discussed, adaptation comes together with innateness---with automatization. 

Notice also that there are useful \textit{operators} in the evolution of networks: The \textit{copying} of nodes along with their connections adds syntactic material to the network from the inside, which serves as a basis for increasing complexity. The \textit{chunking} of nodes and the \textit{severing of connections} between nodes allows nodes to separate from their previous context and join new contexts gradually. 

One thing that is important about this section, and that will become clearer later, is the sense of an Archimedes-screw--like operation of network-level evolution. An Archimedes screw is a helical surface wrapped around a central shaft inside a pipe that is designed to carry water up from one side of the pipe to another as the screw rotates. Each point rotates at its own level, yet due to that rotation, water flows up. Likewise, in network-level evolution, when genetic interaction is replaced by a gene in the course of evolution, or when a behavioral sequence with environmental triggers is replaced by an instinct, there is a sense of a transfer of meaning from phenotype to genotype---from higher to lower levels of organization---despite the fact that materialistic changes like movements of genes are confined to their respective levels (the phenotype does not actually become a genotype). This will help us replace the notion of novelty from a local genetic accident with the idea that novelty arises at the system level and is then crystallized in an evolutionary process based on mutational operators working under natural selection. It also addresses, from an unexpected direction, the fundamental question articulated by Levin of how the different scales of biological organization are connected \cite{Levin1992}. As Levin wrote: ``change is taking place on many scales at the same time, and... it is the interaction among phenomena on different scales that must occupy our attention'' \cite{Levin1992}.

\subsection{The gradual evolution of innateness of alternative splicing patterns results in exon shuffling} 

As an example, the above bears on the evolution of chimeric genes. Traditional discussions on the evolution of chimeric genes seem to assume that they arise by sudden fortuitous events. In contrast, I argue that, as further molecular evolutionary details are uncovered, we will see that such genes are generated by a gradual process. The difference between these views is striking in the case of the evolution of alternative splicing patterns, and here, it brings together various aspects of the present view. 

``Exon shuffling'' refers to the fact that homologous exons can appear in different genetic contexts in different species or even the same species. ``Alternative splicing'' refers to the fact that, in eukaryotes, multiple products can be generated from different combinations of exons, whether the exons are taken from nearby as in the case of cis-splicing, or from different loci as in the case of trans-splicing. The former implies a process in evolutionary time. The latter is a process in developmental time. Now, we know that there are cases where the same exons are being trans-spliced in one species or strain but cis-spliced in another \cite{KongEtal2015}, such as the exons of the separate \textit{eri-6} and \textit{eri-7} in \textit{C. elegans} strain N2 and their fused homologs in \textit{C. briggsae} and in other strains of \textit{C. elegans} \cite{FischerEtal2008}. Likewise, we know that some functions are achieved by multiple single-module proteins 
in one species but by a single, multi-module protein in another, where the genetic sequences encoding these modules are fused \cite{GraurLi2000}. For example, the activities required for the synthesis of fatty acids from acetyl-CoA are carried on by discrete monofunctional proteins in most bacteria, and are encoded by two unlinked genes in fungi \cite{ChiralaEtal1987,MohamedEtal1988} and by a single multi-exon gene in animals \cite{AmyEtal1992} (see \cite{GraurLi2000}). While a connection between exon shuffling and alternative splicing was suggested as soon as the latter was discovered \cite{Gilbert1978}, I offer to sharpen the nature of this connection as follows: exon shuffling is the \textit{gradually evolved innate state} of alternative splicing. Namely, what is constructed in developmental time is \textit{gradually} replaced in evolutionary time with new innate elements and a new developmental construction. Specifically, when two exons previously spliced together at the RNA level are now fused at the DNA level, it is a case where a process in developmental time---a splicing pattern affected by various factors---has become an innate object---a gene fusion, emancipated from the influence of those factors. 

Accidental mutation and natural selection are not suitable for explaining this gradual evolution of innateness of an alternative splicing pattern because it is a long term process that requires multiple changes that interact with each other, each of which is hard to justify by a short-term adaptive value. However, it can occur by mutational mechanisms operating under selection, as discussed in section \ref{trimcypfusion} and in \cite{Livnat2013}. One  
may hypothesize that alleles evolving at multiple loci gradually change the regulation of the alternative splicing pattern in the focal gene as well as in other, coevolving genes. Genetic information from these loci can then be gradually collected by non-random mutation \cite{Livnat2013}, setting the new genetic sequences as well as the new alternative splicing patterns that we see today. In other words, many mutation-writing events, in each of many individuals, in each of many generations, under natural selection, gradually pave the way for network evolution at the gene level. Evolution is a process where many interacting changes happen in parallel over long periods of time \cite{Livnat2013}.  

Two noteworthy precedents to the above are these. First, Stone and Schwartz hypothesized that separate genes whose products first aggregated in the cytosol to form a functioning enzyme could later become fused at the DNA level \cite{StoneSchwartz1990}. They suggested, as an example, that the different lobes of an enzyme such as glyceraldehyde-3-phosphate dehydrogenase may have come from separate genes far in the past, before those genes became genetically fused; and that this could also explain the existence of a family of dehydrogenases, each of which has fused the same gene encoding the NAD binding protein with differently mutated copies of the gene encoding the substrate binding domain. Second, West-Eberhard \cite{West-Eberhard2003} predicted that the connection between evolution and development will be found in the connection between exon shuffling and alternative splicing and in other phenomena; and that somehow what undergoes genetic change during development is also more likely to undergo evolutionary change \cite{West-Eberhard2003}. 
In this paper, I agree with the above and add that the gene-fusion case is merely an example of a more general principle, where meaning is absorbed from context by the gradual change of strength of connections between nodes in a network. 

\subsection{Further insights from the evolution of language}
\label{evolutionoflanguage}

In developing his ideas on evolution, Darwin drew inspiration, among else, from the evolution of language. In \textit{The Descent of Man}, he wrote: ``The formation of different languages and of distinct species, and the proofs that both have been developed through a gradual process, are curiously the same... We find in distinct languages striking homologies due to community of descent, and analogies due to a similar process of formation... We have in both cases the reduplication of parts, the effects of long-continued use, and so forth'' \cite[pp.59-60]{Darwin1871}. Had Darwin known what we know today about the evolution of language and molecular evolution, he would have been able to take his analogy further, and show that principles analogous to those proposed above are essential not only for biological evolution but also for the evolution of language. 

Reminiscent of the ubiquity of cooption in biology, in the course of the evolution of language, words change their meanings as well as adopt multiple meanings. For example, words for ``sharp'' in different languages are related by descent to words for ``tooth'' or ``shard'' of clay, among else; and third person pronouns like ``he'' or ``she'' across different languages are generally related to pointing words for distant objects \cite{Deutscher2010}. The change of meaning is pervasive and the principle of cooption appears to account essentially for all of language \cite{Deutscher2010}. 

Furthermore, the meanings of words generally change gradually, as the following example by linguist Guy Deutscher demonstrates \cite{Deutscher2010}. The word pair ``going to,'' in general and specifically in the shorthand form ``going [to some place in order] to [do something],''  originally meant movement. Gradually, the movement meaning was relegated to the background, while the implication that something was soon about to happen has come to the fore, until ``going to'' has become a future marker, independent of movement  \cite{Deutscher2010}. For example, a sentence from the mid 1400s tells of a travel to some place: ``As they were goynge to bringe hym there.'' A later example reads: ``was goyng to be brought into helle,'' where the passive form ``to be brought'' begins to shift the focus to the temporal realm \cite{Deutscher2010}. Finally, after further such changes, an example from 1642 spoken by King Charles I shows the phrase to mean specifically that something was soon going to happen, without any implication of travel by the subject: "My Magazine [arms] is going to be taken from Me". At that point it was recognized by a linguist as a future marker \cite{Deutscher2010}. 

Note that it was not a sudden change in the words themselves that gave rise to the future marker, but rather a gradual change of context of usage: the more people used the word-pair to emphasize that an activity was soon to take place, the more it has come to be conceptualized in this new meaning. The novelty arose at the system level. Note that there was a hint of the final meaning already in the beginning---when we go somewhere in order to do something, it implies that we will be doing it soon. This meaning was sharpened and gradually released from the previous usage, leading at the end to an abstract concept that applies more broadly than before---to inanimate as well as animate objects. 

Note also that in this fusion of ``going to,'' ``going'' and ``to'' are in some sense duplicates of ``going'' in the original sense of movement, as in ``going to the store,'' and of the ``to'' that is in ``in order to,'' respectively---the latter are the source copies. In fact, in the slang word ``gonna,'' the two words have actually fused in the sense that the space between the words as well as some sounds have dropped. But it is important to notice that an essential part of the fusion had already happen before these local changes, which demonstrates that we must attend to the gradually changing context of usage as leading the process. 

Indeed, not only do new words commonly arise from fusions, they often start with a metaphor that, in the course of the evolution of language, gradually becomes routinized with its own stand-alone meaning. For example, the Old English ``hlaf weard'' (loaf warden; i.e., bread keeper) has gone through the stages of ``hlaford,'' ``laferd,'' ``lowerd,'' finally providing us the abstract ``lord'' \cite{Deutscher2010}. The Latin de-caedere, or ``cut off,'' has evolved into ``decide'' \cite{Deutscher2010}. (Note the metaphor between the literal meaning of ``bread'' and ``keeper'' on the one hand, and ``lord'' on the other, for example.) Indeed, metaphor is a metaphor of itself, because it literally means carry across from one context to another (meta: across; phor: carry) \cite{Deutscher2010}, which is our topic---cooption. Furthermore, it is a common occurrence that when two words are used frequently and obligatorily together in an emerging context, their independent existence in that context becomes irrelevant, and they are shortened and fused into one word---which is reminiscent of the \textit{TRIM5-CypA} fusion mechanism. 

The above provides also an analogy to innateness. In the beginning of the use of a pair of words that are to become a word fusion, the pair is constructed using the ability of speakers to put together previously learned words into combinations with their own meanings, and is understood using the ability of listeners to analyze combinations in terms of the words they are made of. But as the two words come to be used more and more frequently together in a certain emerging context, they 
come to be perceived less and less as a constructed phrase and more and more as a word in its own sake. That is, increasingly the new fusion is learned by children directly from the context of its usage at the same time as other words are learned, rather than being constructed figuratively during speech.  Eventually it is hanging by its context alone. It is no longer constructed from units more elementary than itself, but is a new elementary unit with its own literal meaning. In this quickening of the construction of the new fusion until it becomes an elementary unit there is a metaphor for the evolution of innateness. 

In summary, the new elements of language are not invented out of thin air. Rather, the source for their creation preexists at the system level. One may say that an essential point about human language is that it allows us to put together words into phrases and sentences that communicate novel meaning; but note also that 
from these phrases and their contexts of usage, new words arise. The vocabulary grows in a manner connected to word usage. And as this vocabulary grows, our ability to express meanings is refined. Whereas previously ``going to'' had the explicit meaning of travel and an implicit meaning of ``soon,'' now we have both, including a clear, separate meaning of ``soon'' that is applicable in new situations. Thus, from the ambiguous that can play multiple roles, come the distinct, diversified and specialized. The process ``starts'' at the system level. 

\subsection{Novelty comes neither from a point nor from DNA ``misspelling''}

Now, gene fusion may be discussed as one topic, and cooption as another. But they are actually two sides of the same coin.  In both cases we see elements or copies thereof leaving their previous context and moving to a new context. But although fusion and cooption are 
parts of the same process, the case of fusion is especially grabbing to the eye, because it shows the creation of a new elementary unit in a manner that traditional theory has not prepared us for. In traditional theory, there is point mutation and presumed novelty from it, and there is gene duplication followed by point mutations in the duplicates \cite{Lynch2007}, but there is no evolutionary process where \textit{a process can become an object}---where a new elementary unit is created from something that previously was an interaction (indeed, this new elementary unit absorbs new meaning from its gradually changing context).

Two remarks are important. First, this manner of creating a new elementary unit requires the existence of a hierarchical structure of organization---a network---where, by a gradual change in the network, such a process can happen. Since this hierarchical structure does exist and is a fundamental aspect of nature, it is an advantage of the present theory that it engages this structure\footnote{In contrast, in traditional theory, genes are often perceived as independent actors, and mutation is perceived as a local genetic accident that brings new phenotypic meaning on its own. Traditional models do not have a representation of the phenotype---of biological structure---and therefore treat genes more as beads on a string than as nodes in a network. They are oblivious to what is happening above the bottom level of the biological hierarchy, and to the possibility that from higher up comes a force that changes something at the bottom level. They simply assume that the bottom level of the hierarchy is in control all on its own of what is happening evolutionarily, by means of random mutation.}. Second, the gradual creation of a new elementary unit from what was previously an interaction is important because it shows us that the \textit{barrier between ``unit'' and ``interaction'' has been broken}. There is no sharp dividing line between elementary units and higher-level interactions. As with the fact that the phrase ``going to'' never needs to become the fusion ``gonna'' in order to become a word for all intents and purposes---a unified concept, automated and regulated as one---and as there is no particular point in time where it suddenly turned from two words into one, so there is no clear line telling us when two exons that interact need to be considered as making up one gene as opposed to belonging to different genes \cite{GersteinEtal2007}. The collapse of the gene concept as a well-defined unit is supportive of this absence of a sharp division between process and object\footnote{We now know that genetic elements we previously thought to participate in ``one'' gene actually form products together with elements that we previously thought to belong exclusively to ``other'' genes, and so the boundaries between genes have been much blurred.} \cite{GersteinEtal2007} and fits with a \textit{gradual} process of gene formation. 

Indeed, the view proposed here is importantly different from the traditional one. Not only does the traditional view focus on object minus context and claim that novelty arises in the object by a local genetic accident that emanates this novelty ``upward'' to the complex system---novelty from a point---but in addition, this point-like change is considered to be an error akin to a ``misspelling.'' If we let genes be words, metaphorically speaking, and let the phenotype be the technology that they describe, then the traditional notion of mutation can be exemplified by misspelling unintentionally the word 
``incubator'' while making all effort to copy the word 
``incubate'' accurately, and thus suddenly getting the idea of inventing an incubator. Whereas in reality, the 
incubator (i.e., technology, or the phenotype) is invented by the use of many concepts described by putting together many words; and in the long-term, the whole complex object that is an incubator might even be given a standard, symbolic name by which this whole has come to be referred to conveniently: the word ``incubator,'' generated by a standard operation of adding the appropriate suffix to a preexisting, useful word. From the view of interaction-based evolution, to say that the ``misspelling'' of genes is the source of biological novelty is to make a mistake in understanding the nature and the role of the bottom level of the genetic interaction hierarchy, similar to saying that the misspelling of words creates technology. 

\subsection{Evolution is a bottomless system}

Considering all the above, we can now describe a main point of this paper. Evolution is a ``bottomless system\footnote{This term, which aptly describes one of the most important points of this paper, was proposed by Nick Pippenger.}.''  One cannot define all words in the dictionary in terms of other words without getting into a circularity. Ultimately, the meaning of words comes from the context of their usage; that is how language is learned and even how it evolves. The genes are similar in this regard. Their meaning comes from their context of usage. They themselves are nodes in a network, in development as well as in evolution. The upshot of this is that the bottom of the hierarchy of biological interactions---the genetic sequence---is not a stable ground upward of which life is built. Mutation is not a local accident that brings innovation all on its own 
as though there is no living network that it needs to connect to. The process of genetic change is a complex one where the connections between nodes in the network become stronger and weaker as they form modules that \textit{absorb meaning from context}. 

With this key, we will begin to replace the source of novelty in evolution. Traditionally, we have been thinking about an accident, disconnected from the living network, as an event that creates new information. This was conceived of as a point-like event, which then emanates the novelty that it brings about to the phenotypic level. I argue instead that novelty arises from network-level change, not from a point. This involves a mutation-writing phenotype that executes network change in a syntactic and evolving fashion \cite{Livnat2013}. 

\section{Simplification and novelty} 
\label{simplification}

For Darwin as well as for Fisher \cite{Fisher1930}, complexity evolved in cases where an increase in it was needed for an increase in fitness. However, the question of why complexity evolves has never been resolved \cite[p. 11]{Wagner2014}. I argue here that simplification under performance pressure leads to both complexity and novelty. 

This section will be entirely devoted to discussing the concepts. Once they are discussed, the numerous empirical examples given in section \ref{innateness} can be understood. 

\subsection{Simplification under performance pressure leads to complexity} 

Several points in the present theory may be organized under the heading of ``simplification,'' each of which comes with its own corresponding increase in complexity. 

\subsubsection{Modularity and simplification}
\begin{itemize}
\item Simplification and modularity are tightly connected concepts. A module serves multiple contexts---in fact it is defined by them---and in the case where one serves the many, as in the case where one explains the many, there is frugality, parsimony, or simplification. 
\item I discussed above the gradual appearance of modules in networks. A key example of the appearance of a module was the fusion of two genetic elements. Here, the developmental process originally putting them together is simplified away in the course of evolution. More generally, the gradual arising of new modules from a previously complex, interconnected mass of nodes is the evolutionary streamlining, or simplification, of development. Elements inside a module are emancipated from the complex influence of elements that are now outside of it and are no longer connected to it. 
\end{itemize}

\subsubsection{Innateness and simplification}
\begin{itemize}
\item As will be shown soon, an extension of the last point is the evolution of innateness, which involves evolved independence from environmental triggers. During evolution, 
an evolving trait can become emancipated from 
complex environmental influence involved in the development of an adaptive ancestral phenotype as a more orderly, simplified and compartmentalized developmental process evolves. Thus, the evolution of innateness involves simplification: what consumed developmental (and sometimes learning) time is simplified away in the course of evolution.
\end{itemize}

Now, the cases of simplification described above come together with an increase in complexity. As argued earlier, due to the duplication of genes, the formation of a new module need not come at the expense of old modules. The increase in the number of modules or elementary units comes together with an increase in the number of interactions between such modules or units, which represents an increase in complexity. Simplification is what we see when we look at the modularization of an  interconnected mass, and complexity is what we see when we look at emerging interactions involving newly formed modules. \textit{Local simplification leads to a global increase in complexity.} 

\subsubsection{The final touch of perfection}
\label{finaltouch}

There are observations that show the development of organs or tissues taking ever straighter paths over evolutionary time \cite{Muller1869}. For example, in cetacean embryos (e.g., whales and dolphins), hind limb buds still appear fleetingly in development and grow to a small size before they are removed \cite{SedmeraEtal1997}. In such cases, it is evident that, over evolutionary time, the developmental process gradually comes to spend less and less time and energy on developing structure that is slated to be superceded by another or to be removed later in development. 

How does it happen that evolution straightens up developmental paths? A neo-Darwinian answer is that the savings of time and energy are directly favored by natural selection, so that a whale that acquires by chance a mutation that reduces the development of the useless bones by even a small amount gains a slight benefit in terms of survival and reproduction, and thus accidental mutations of this sort are passed on preferentially. We must ask, however, whether it is reasonable that a slight straightening of the developmental path of useless, internal small bones is truly enough to make such an impact on differential survival and reproduction that would be noticeable, when presumably many and much more important other individual differences contribute to differential success. Indeed, the problem of the obliteration of rudimentary organs is a very old one \cite{Darwin1859}, and was discussed by Weismann hand in hand with that of the final touch of perfection---how adaptations become perfected beyond what seems to be possible by traditional means \cite{Weismann1902}. Darwin himself \textit{agreed} that it was not possible to explain the removal of rudimentary organs as an outcome of natural selection alone based on minute economic considerations \cite{Darwin1876nonote}. Indeed, in light of the sections to follow on innateness, it is remarkable that he held steadfastly to the Lamarckian ``laws of use and disuse'' to explain them. And if Darwin is not neo-Darwinian enough to defend the latter, then one may consider the father of neo-Darwinism, August Weismann---who is responsible for the rejection of Lamarckism: How did he explain the final touch of perfection? By suggesting a principle of ``momentum'' or ``inertia,'' where a mutation in a certain direction will be followed by others in the same direction, so that noticeable, selected improvements of economy will be followed up by minute, unselected ones \cite{Weismann1902}---a point which is completely outside of the view based on random mutation and natural selection, a view which traces its ideological origins to Weismann. It seems that no serious explanation was found for these phenomena within neo-Darwinism, and indeed those who were supposed to be the two greatest pillars of it went to great lengths to look for alternative explanations. 

The theory proposed here tackles this old, unresolved problem head on. It argues that it is not accidental mutation, but simplification, that explains the final touch of perfection, both in the complete obliteration of a trait and in the crystallization of adaptation (see section \ref{backwardwalking}). In addition, Weismann's idea of inertia is not beyond the pale for a theory where the writing of mutations has evolved under the influence of past selection. 

Now, notice again the connection between simplification and complexity: the intriguing straightening of developmental paths demonstrated by the unexplained old observations is tied to the ``final touch of perfection''---a honing in on 
an optimum in the evolution of a complex adaptation. 

\subsubsection{Convergence, simplification and complexity}

According to \cite{Livnat2013}, the writing of mutations over the generations combines information from different loci and from different individuals that succeeded in survival and reproduction. Alleles at different loci concomitantly spreading in the population do not each bring an independent piece of the phenotype to all individuals, but rather interact with each other. Thus, an adaptation evolves at the level of the population as a whole, at the same time as it 
becomes more genetically stable \cite{Livnat2013}. This process slowly \textit{gives rise to the true, common reason for success shared by individuals}, as the initially many and highly variable ways by which different individuals approximate the adaptation only roughly at first are gradually superseded by an adaptation, uniform across individuals (see nest-digging by sand wasps, discussed in section \ref{Evans} and in \cite{Livnat2013}). We may now note that in this replacement of many by one---of the different rough approximations by one uniform adaptation---there is simplification. At the same time, this one that replaces the many is a complex adaptation---a point of optimality. Therefore, simplification and complexity again come together: the complexity that is in the different ways of approaching an adaptation has been converted into the complexity of the adaptation itself. 

\subsubsection{Simplification and complexity: summary}

We have seen that each of the above connections to simplification comes together with an increase in complexity. Could simplification under performance pressure (e.g., under selection) be the cause of the evolution of complexity? This question is best answered together with another, related question, discussed next: What is the source of novelty in evolution? 

\subsection{The problem of novelty}
\label{problemofnovelty}

Lamarckian or ``adaptive'' mutation has been the only alternative so far to accidental mutation\footnote{As noted, the evolvability approach implicitly or explicitly relies on accidental mutation as the ultimate cause of heritable novelty.}, but it has fundamental problems. First, it does not apply to multicellulars: there is no intra-organismal mechanism that senses that the hawk needs sharper vision and then makes the genetic changes in the germ cells needed to bring about that phenotypic change. Second, hypothetically speaking, even if there were mutational mechanisms that knew what would have been favored by natural selection in a particular organism at a particular point in time and how to produce it, this would not have solved the problem of how novelty arises, because the novelty would have been in how such supposed mechanisms acquired that particular knowledge to begin with. Indeed, it is easy to erroneously think that, 
if there is knowledge of the thing to be produced, there is no novelty, and if it is to be produced without knowledge, it must be produced by accident. Thus, we can understand the immense attraction of accidental mutation from a traditional perspective: First, it requires no impossible mechanism transferring knowledge from the macroscale to the genotype. Second, accident has no preconceptions, and it seems to have been believed that it can invent almost anything---that it can produce novelty. 

As articulated in \cite{Livnat2013}, it is a key property of interaction-based evolution that the non-random mutation that it proposes does not circumvent, but rather works together with, natural selection, and is strictly non-Lamarckian. This removes the first reason to hold on to accidental mutation. However, if the mutational mechanisms are not ``aware'' of the environment, how could mutation be anything other than accidental? How could the ultimate source of novelty in evolution be anything other than random mutation?   

\subsection{Simplification under performance pressure leads to novelty}
\label{simpunderperf}

To try to answer this question, let us allow ourselves to step outside of evolution and look at how novelty arises in other creative processes. 

Consider the development of scientific theories. It has two fundamental principles. First, theories need to fit the data---they need to \textit{perform}. Second, they must be parsimonious. When we take disconnected facts and find a theory that explains them all in one, we create a more parsimonious picture of reality than existed before. It is a fortunate fact of nature that when we do so we often obtain a model of reality that will hold better when new and unexpected data later arises and that will lead to findings not previously expected. 

A well known example of the use of parsimony in science is the Copernican revolution---the placing of the sun instead of the earth at the center of the solar system. Copernicus proposed this model not because it allowed him to make better predictions of the movements of the planets, but because it was simpler on an essential point \cite{Singh2004}. This simpler model paved the way to future science, generally fitting with major later findings by Kepler and Galileo, like the phases of Venus.  

From this and many other examples we see that the pursuit of parsimony does not merely provide elegance \textit{per se}. Parsimony \textit{expectedly brings the unexpected}---useful things that were not initially predicted and were not the goal of the work, yet commonly appear \textit{as a result of work}. By simplifying under performance pressure we do not act randomly. Rather, we put work in, and get novelty out: a new, useful prediction or connection emerges that was not originally expected.  Thus, it is not the case that either one knows one's goal and there is no novelty in getting there, or one does not know it and the only way to get there is by accident. Rather, there is a third way to novelty. 

Several important comments follow. First, we need not explain why simplification under performance pressure leads to novel, useful things in science that were not directly sought. For now, we may simply take it as a grand fact. 

Second, importantly, this simplification does not make science as a whole simpler but rather more complex. As new theories connect between previously unconnected facts, new predictions and new questions arise. The more knowns there are, the more they interact and expand our ability to ask yet new questions. Thus, I argue that simplification under performance pressure leads to both novelty and complexity. 

Third, simplification and performance function together. As statisticians or investigators in machine learning know, it is useless to make a model that predicts a given set of data points perfectly if the model is overly complicated, as it is useless to set up a model that is very simple but has nothing to do with the data. A balance must be maintained between fit to data and model elegance, and to maintain it is an art. 

Indeed, the desires for simplicity and for performance are \textit{conflicting:} at the time when Galileo originally favored the Copernican over the Ptolemaic system, he did it \textit{despite} the fact that the former fit the data a \textit{little} worse, and because of the fact that it was \textit{much} more parsimonious. Indeed, later scientific research showed that the more parsimonious model was far more \textit{improvable}.    

The development of mathematics gives us a similar picture. It happened once and again in history that pure mathematicians working on the principles of aesthetics or parsimony have produced things that years later were found to have unexpected utilitarian value \cite{Wigner1960,Hamming1980,Byers2010}. Indeed, the power of operations other than the test of performance in the growth of mathematical and scientific knowledge has been amply demonstrated. We see it in simplification or parsimony, elegance or aesthetics, symmetry, pattern completion and analogy \cite{Wigner1960,Hamming1980,Byers2010}. I use the word ``simplification'' in a very broad sense to refer to all these variants and the creative force they represent. Note also that in both mathematics and science, 
we operate with a network of concepts. We connect between ideas to create a fuzzy, new idea, distill a fuzzy new idea to its essence, and pursue the consequences of a distilled idea to new connections (Christos Papadimitriou and Umesh Vazirani, personal communications). Thus, novelty arises from the network, not from random, point-like changes. This network change is driven by both simplification and performance, and we can see that it leads to complexity, novelty and improvement. 

The evolution of technology is also illustrative. What is simple appears in many different technologies. The concept of a disc appears in the potter's wheel, in wheels for transportation, in a round table, and in the cross section of a tree trunk. The concept of a sharp edge appears in a stone tool, a peg, and even a shingle roof. Once we generate a functional but elegant object in one context, it is going to have the inherent capacity of working well in future, different contexts. 

I argue here that, also in biological evolution, simplification and performance pressure, and not accidental mutation and performance pressure, drive complexity, novelty and advancement. This new theory has an advantage over the previous one. When we rely on simplification under performance pressure, we rely on something that we can see to be central to other creative processes. 

A key aspect of simplification is that it allows us to circumvent the problem posed earlier: how mutation can do anything useful, how it can be anything besides accidental, without ``awareness'' of the environment and the macroscale phenotype. The solution is that biochemical work that simplifies local connections in the genetic network requires no knowledge of the macroscale phenotype and the environment, and can take place in the germ cells. That is, while local simplification and gene duplication operations take place in the germ cells, natural selection evaluates the organism as a complex whole, and together these two forces lead to novelty. This allows us to replace the concept of accidental mutation with a concept of \textit{non-accidental mutation that is useful yet not Lamarckian}, and thus to replace the traditional notion of random mutation as the ultimate source of novelty in evolution.  

\subsection{Where simplification and performance pressure happen}
\label{wheresimplificationhappens}

In addition to simplification pressure at the genetic level and performance pressure at the organismal level, each of the two may have, at its own level, the other on the other side of the coin. For example, in an ecological community, each species is pressing to produce more of itself and at the same time is undoing the growth of others, thus pressing to simplify the ecological network. The same could be said of a gene that comes to replace another in the course of evolution by usurping the other's role, a phenomenon called ``genetic piracy'' by Roth \cite{Roth1988} (see also \cite{Wagner2014}). The ecological example above clarifies that the implementation of simplification can be as basic and follow as naturally from the situation as differential survival. In fact, here they are two sides of the same coin: inasmuch as the making more of one entity means making less of another, the performance of any one entity puts simplification pressure on the network, and this principle may apply both to the ecological network and to the genetic network. It is also noteworthy in this regard that a gene that is extensively used (performs well) and therefore highly expressed may, due to mutational mechanisms, be more likely to be duplicated. This will be relevant in section \ref{mutationalmechanisms}. 

\section{The problem of innateness}
\label{innateness}

It is time to substantiate the ideas proposed in this paper with many examples from the phenotypic level. This section will do so with the help of empirical observations relating to one of the oldest and most mysterious problems in evolution---the problem of the evolution of innateness. Although the observations to be discussed are each known and available in the literature, here I will argue for their fundamental importance in evolution through a connection with interaction-based evolution. I will first cover innateness from multiple angles in sections \ref{preex}--\ref{unitesdifferentaspects}, and then discuss the emergence of novelty in detail (section \ref{backwardwalking}). Readers interested in the molecular level may note that it will be revisited in section \ref{newview}.

\subsection{The problem of the preexistence of high-level mechanisms}
\label{preex}
 
The ability of pointer dogs to point at the prey in a statuesque manner (among other abilities) is to a large degree innate \cite{Arkwright1902}. How did this instinct evolve? To argue that a sequence of random mutations of small effects has built up the behavior from scratch such that it has always been instinctive and never learned is unappealing: Would breeders have recognized slight inborn tendencies to point at the beginning of the evolutionary process involved and,  without regard for the outcome of any training, base their artificial selection on these differences? And if training was important in the evolution of pointing, the highly evolved abilities of the animal to learn would have masked out presumed mutations of slight effect for an \textit{independently developed} instinct. All would be more understandable if we consider that a trait that previously required learning through reward and/or punishment has become emancipated in the course of evolution from these external cues. 

Consider the evolution of migration. In an instinctive and automatic manner, a young common cuckoo (\textit{Cuculus canorus}) takes off in the fall from its breeding grounds in Scandinavia, flies thousands of miles to its wintering site in Central Africa and then returns in the spring \cite{WillemoesEtal2014}. How did this complex suite of instincts get started in evolution? Both Darwin \cite{Romanes1883} and Wallace \cite{Wallace1874} hypothesized that the breeding and wintering grounds gradually became separated and the distance between them increased; that originally, the animals were tracking seasonal changes in resources over short distances as a direct response to the environment; and that in time this behavior has become habitual and instinctive \cite{Romanes1883,Wallace1874} (see also \cite{Woodbury1941}). To assume that the migratory instinct evolved afresh, independently of the behavior that came before it, brings up the same problem as in the case of the pointer dogs: the pre-existence of an evolved, general-level mechanism (in this case, the brain) that is able to respond adaptively to environmental changes and was presumably involved in the original phenotype.

In an experiment designed to capture the evolution of innateness \cite{Waddington1953} (see also \cite{Waddington1956a,Waddington1956b,Waddington1959,Bateman1959a,Bateman1959b}),
Waddington took \textit{Drosophila melanogaster} flies and exposed their pupae to a heat shock. As a result, a fair number of the flies that developed showed a particular vein pattern on their wings---an absence of or a gap in the posterior crossvein and sometimes the anterior one too---called ``crossveinless.'' He then bred the crossveinless flies to form the next generation of the experiment and repeated this procedure of heat shock and selective breeding over the generations. As a result, the percentage of crossveinless flies increased over the generations and, beginning at generation 14, a small percentage of flies started showing the new vein pattern without exposure to heat shock, that is, innately\footnote{In order to observe this, the experimenters took at each generation a certain sample of flies and raised them without heat shock.} \cite{Waddington1953}. The fact that this trait became innate, \textit{when no selection for such innateness had been performed}, is an intriguing experimental outcome called ``genetic assimilation'' \cite{Waddington1953}. 

To explain genetic assimilation, Stern \cite{Stern1958} (see also \cite{Falconer1960}) proposed a model based on traditional principles. The model assumes the preexistence of alleles that make independent contributions toward a certain sum, such that if the sum surpasses a certain threshold, the trait of interest is exhibited. Furthermore it makes certain assumptions about the initial frequencies of alleles and the normal conditions and experimental conditions thresholds that make it so that, prior to selection, the trait of interest (e.g., crossveinless) is exhibited in practice only under experimental conditions (e.g., heat shock), whereas post selection it is exhibited under both experimental and normal conditions, and thus the trait can be said to have become innate. However, despite its mathematical crispness, taken literally, this model means that every trait that is to become innate has its own set of additive alleles that preexist and provide the potential for that trait to become innate as is. That is, there are additive alleles that, if they surpass a threshold, build a brain that points, and there are different sets of additive alleles lying dormant in birds for every possible migration route, such that each set builds a brain for a particular route if it surpasses a certain threshold\footnote{In fact, once we assume that Stern's model taken at face value is the relevant method of explanation, it would have been easier to assume that complex instincts in nature evolve afresh, without relation to a preexisting behavior modulated by a brain and modified by the environment, because the model does not describe a world where such a relation is biologically reasonable---it requires a brain that affects independently threshold expediently assumed for each particular trait that is to become innate, each with its own expediently assumed set of additive alleles.}. Indeed, Waddington himself rejected this model \cite{Waddington1958,Waddington1961}, because it did not apply to the complex cases that motivated the problem. Here, I will provide another explanation for innateness based on network evolution.

\subsection{Modularity and innateness are caused by simplification} 
\label{modularityandinnateness}

As Waddington alluded to \cite{Waddington1942}, when an emerging module is released from the influence of an element inside the organism, the result is seen as modularization; and when it is released from the influence of an environmental factor, the result is seen as the evolution of innateness. 
Earlier I argued that simplification is connected to modularity and innateness: the formation of modules streamlines the developmental process and involves emancipation of an emerging module from complex influences, both internal and external. Indeed, simplification leads to modularity and innateness. 
 
Approaching the topic of innateness equipped with the theory of gradual network change presented here, it is useful to distinguish between two important phenomena that I will call ``emancipation'' and ``acceleration.'' Emancipation refers to the fact that nodes (modules or elements) can be copied and the connections between nodes can gradually evolve such that a node can be subjected to a different regulation than that of its source copy. Acceleration refers to the idea that the coming together of nodes under one control simplifies development locally while absorbing novel phenotypic meaning from the changing context. Both these aspects of network level evolution, discussed in section \ref{verbalmodel}, will be clarified with the help of examples, and both figure into the explanation of innateness to be given in the following sections. 

\subsection{The evolution of innateness is more common than we realize because the innate, derived phenotype is usually not identical with its non-innate, ancestral source}

In an idealized view of the crossveinless experiment, we can think of the crossveinless trait as qualitative (present or absent) and assume that it is the same in the beginning of the experiment as it is at the end. The only thing that evolves under this assumption is the propensity to produce it. In this case, we may simply use the word ``emancipation'' to describe what happens to the crossveinless trait when it comes to appear without the environmental trigger. But crossveinless is 
an extreme, chosen for its simplicity. In nature, when the evolution of innateness or emancipation takes place, the trait that is to become innate also evolves at the same time. For example, in cases of ritualization, a non-signaling behavior is gradually released from its context and becomes used as a signal (e.g., an egg-fanning movement becomes a showing-the-nest signal \cite{Tinbergen1951}; see section \ref{elementsofnetworkevolution})  \cite{Huxley1914,Whitman1919,Armstrong1950,Tinbergen1952}.
As Tinbergen noted, those ritualized traits that are emancipated are usually traits that have already changed much from their original form; and we would not have been able to make a connection between the signal and its origin if it were not for the fact that, at least in some cases, there happened to be a transitional series betraying the connection between the two, such as in the threat posture of the Manchurian crane (\textit{Grus japonensis})
\cite{Lorenz1935,Tinbergen1952}. In other words, there exists a continuum of differences between the non-innate ancestral and the innate derived traits, that ranges from no difference, to a great difference that obscures the connection between the ancestral and the derived; and cases at the former end of the spectrum are rare. 

I argue that this is precisely the problem with observing innateness. Darwin and other early naturalists believed that what is habitually performed due to environmental triggers over the generations gradually impresses itself on the hereditary constitution of the species and becomes innate and emancipated from the environment, and that this is explained by the laws of use and disuse, or Lamarckism \cite{Romanes1883, Darwin1859}. I argue that such automatization happens in general but is often hard to see because of the difference between the ancestral and derived traits past the point of cooption, and that it is network-level evolution and not Lamarckism that is responsible for it.   

A spectrum of differences between the new innate and the old non-innate is predicted by interaction-based evolution. If we do not recognize this spectrum, we are liable to notice only the easily visible cases at one end of 
it and then falsely argue that because of their rarity we can continue business as usual. However, it is better to 
recognize that what we easily see of this spectrum is its extreme, which is the tip of the iceberg. 

\subsection{Evolution of the whole as a whole in innateness} 

The fact that a trait changes as it becomes innate allows us to examine the evolution of a complex whole, while involving not only emancipation but also welding and acceleration.  

While I have been using the term ``innate'' without qualification so far, it is useful to note that there is no strict separation between the innate and the non-innate. Learning itself is enabled by instinct \cite{Lorenz1965,GouldMarler1986}. No trait develops in a manner that is independent of the innate nature of the organism, and no trait develops entirely independently of the environment, when the latter is broadly construed \cite{Lehrman1953c}. Therefore, rather than speaking of ``innate'' and ``non-innate,'' we realize that there is a continuum between things developed more directly and quickly (``innate'') and things that require more unfolding that involves more interactions with the environment.

When we consider this continuum as it applies to a given organism, we should consider that it evolves as a whole---the process of development evolves as a whole. Then, we can bring the ideas of gradual network evolution to bear on it. 
I argue here that the evolution of innateness arises 
as a result of the ``chunking'' or modularization of a previously complex part of the network---what were previously independent elements each under a different control have now become simplified or combined into a singe unit. Although it may seem that this simplification accelerates development in the course of evolution toward the final trait, in general, development is not accelerating toward the final trait as it was before, but rather toward what that trait has in the meantime changed into, and therefore nothing is being accelerated strictly speaking. Therefore, it is the signature of the previously less innate that we generally see in the current more innate, rather than a direct facsimile. There is no Lamarckian transmission that takes a developed or a learned trait and makes it innate. As argued in section \ref{verbalmodel}, to use a metaphor, in an Archimedes screw, water is moved along the shaft even though each point in the screw only rotates at its own level. So in evolution, the non-innate does not itself become innate---the phenotypic does not become genotypic---but rather evolutionary action at each level of biological organization remains at that level, while accelerating development.

We know that the adult form influences the evolution of the earlier stages of development: there is selection on the adult form, and therefore there is selection on earlier stages of development to lead to a well-performing adult. When considered from a traditional standpoint, this trivial point turns into a problem because \textit{development is a complex process}, and traditional evolutionary theory is not equipped to deal with a complex process. Traditional evolutionary theory cannot conceive of a complex evolutionary change that modifies the developmental process as a complex whole: it does not have a sense of acceleration\footnote{but see the lively debate in the 19\textsuperscript{th} century on it \cite{Gould1977}.} or an emphasis on emancipation, and therefore when a trait appears earlier in development that seems to relate to one that used to come later in development (e.g., innate migration relates to earlier, learned migration), it absurdly has to invoke an evolution of that trait afresh, absent any connection to that which it obviously relates to. While Gould attempted to address this problem, he did so by breaking the whole again into parts and arguing that the timing of appearance of one part or a developmental process in and of itself can be moved earlier or later in development \cite{Gould1977}, which is a very limited explanation that does not address the range of phenomena discussed here. 

I have presented, in contrast, a view of the evolution of the whole as a whole. Instead of the 
evolution ``afresh'' idea that arises from a traditional perspective, this view raises the notion of acceleration as described. The quicker arriving at an evolving developmental outcome has the appearance of the evolution of innateness, thus involving interaction-based, network-level evolution in innateness. We have seen this in the case of the TRIM5-CypA fusion and the evolution of alternative splicing patterns at the molecular level, and will now see it in many examples at the phenotypic level. 

\subsection{Pointing in pointer dogs as an example of the importance of the complex whole in innateness} 
	
Pointing in pointer dogs will serve as an example of the importance of the evolution of the whole as a whole in innateness. I propose that selection has operated on the outcome of the training, favoring hunting dogs whose behavior following training was more pleasing to their owners, specifically in stopping upon discovery of the prey instead of chasing it further. However, since innate tendencies guide the learning, this selection has operated indirectly on innate tendencies, favoring dogs with the right set of innate tendencies that were more naturally inclined to learn the right behavior. In particular, I hypothesize that there exists among animals a widespread, natural tendency to heed sudden changes; that in pointer dogs, this tendency has been strengthened in the course of evolution, in particular by heightening nervousness; and that other instincts have become at the same time adjusted to direct it productively, helping the dogs to learn to pause and freeze at the sight of prey.
Over the generations, the learning task has become more and more natural to the dogs, and the amount of learning required has decreased, until today, the dogs require only minimal training, and they sometimes point at objects innately without any learning, as Darwin observed \cite{Darwin1909}. Thus, selection for an improved outcome of the learning was accompanied by an acceleration of the learning and, ultimately, innateness. This hypothesis, presented here in its specific form that applies to pointer dogs, already has the advantage that it explains also the nervousness syndrome that often afflicts these dogs: heightened nervousness helps them heed sudden changes, and when not properly compensated for results in the nervousness syndrome. Note that, while Grandin and Deesing \cite{GrandinDeesing1998} argued before that pointing relates to nervousness, their discussion seems to assume that pointing and nervousness are traits with separate genetic causes that happen to be connected through genetic linkage, which suggests a spurious connection. 
In contrast, the hypothesis proposed here connects pointing and nervousness at a deep level, with the help of the holistic view of interaction-based evolution. There are no genes dedicated to pointing \textit{per se}: pointing is a system-level phenomenon that emerges from a 
suite of interacting instincts and learning. 

In an exceptionally inspiring chapter, Papaj has already argued that what is first learned can come over evolutionary time to be learned more quickly until it eventually becomes innate \cite{Papaj1993}. In this respect, his hypothesis is similar to the above. However, lacking the ideas of interaction-based evolution, and treating instinct and learning as separate elements, he tried to create a model of a traditional kind, and admitted that the model failed to provide an explanation, because the evolution of innateness came out of an artificiality built into it \cite{Papaj1993}. 
In contrast, the hypothesis presented here allows us to preserve Papaj's intuition but in a natural way: 
it holds that selection has affected interacting instincts that guide the complex process of development and learning through a process of network-based evolution, and network evolution involves acceleration and emancipation---an increase in the innate abilities. In addition, interaction-based evolution also explains why innateness and stereotypy are deeply intertwined (see sections \ref{degreesoffixedness}, \ref{unitesdifferentaspects}), which Papaj's model does not \cite{Papaj1993}. 

To reiterate, I propose that the process of evolution toward a better outcome of the learning 
leads at the same time to a quickening of the learning and that ultimately, a new innate trait appears, because what enables the organism to reach a better outcome through learning is that it is naturally inclined in 
the right direction. The organism ``gets it'' more, naturally and inherently, because underlying interacting instincts are being shaped. Thus, \textit{improvement comes together with innateness.}  

Importantly, the evolutionary process that shapes the network of underlying instincts can be seen as uncovering better ``principles'' that guide the learning (and more generally, development)---emerging underlying elements that organize a preexisting complex more simply. Consistent with section \ref{simplification} and with later sections (see section \ref{backwardwalking}), viewing things in terms of such principles leads us to a new prediction regarding novelty: the evolution of the new innate and the new and improved adult form will come together with the production of new, beneficial things \textit{that were not selected for in and of themselves but arose as corollaries or windfalls} of figuring out the right principles. Improvement, innateness, \textit{and generalization}---or the emergence of useful novelty not selected for---come together. As an example, backing in pointer dogs\footnote{``Backing'' refers to the fact that these dogs copy the posture of another dog who is on point; even this behavior, which is of use to the hunters, is strongly innate.} may have evolved as a ``corollary''---an unintended but desirable outcome. 

\subsection{Elements of network evolution: chunking, emancipation and rigidification}
\label{elementsofnetworkevolution}

As noted, ritualization is an evolutionary process that occurs when a behavioral element is gradually emancipated from its original use as it becomes coopted for use as a signal in the course of evolution (see, among else, \cite{Huxley1914,Whitman1919,Armstrong1950,Tinbergen1952} and further references below). For example, when a bird is about to hop or take flight, it bends its legs, lowers its breast, raises its hind parts and sometimes its tail, folds its neck and brings its head back almost to the shoulders, while slightly expanding its wings, so that the whole body is like a tight spring ready to be released for jumping, at which instant the legs straighten, the breast and hind parts line up with the direction of the jump, and the neck is stretched forward. In an ethology classic, Daanje \cite{Daanje1950} argued that, from this movement, various signals have evolved. For example, when the male turkey displays to the female, it raises the hindparts a bit, raises and spreads its tail, folds its neck and brings its head back almost to the raised back feathers, and partly spreads its wings downwards. This posture imitates that of the jump in several elements, except that the legs are not bent, the tail and wing movements are exaggerated, and the posture is kept frozen for a while \cite{Daanje1950}. Thus, a behavioral pattern which is widespread taxonomically and which originally had a mechanical function has evolved into a signal that is expressed in new contexts independently of the context of expression of the ancestral trait.

Another example of ritualization is the way that the male three-spined stickleback (\textit{Gasterosteus aculeatus}) shows the nest entrance to the female. According to Tinbergen, this movement was derived from the egg fanning movement \cite{Tinbergen1951}, which again demonstrates a shift from one context to another. 

As we have just seen, ritualization requires \textit{emancipation} from one context and \textit{cooption} to another. Critically, these are operations of network evolution. The gradual release of an element from one context concomitant with the subjecting of it to another context involves two aspects of the organism at once and is inherently an interactive operation, well-described by modules moving in a network. It is \textit{not} well described by the traditional notion of evolution as a process affecting ``one thing at a time.'' 

Baerends's work on nest building, egg laying and offspring provisioning in the digger wasp \textit{Ammophila adriaansei} (\textit{campestris}) \cite{Baerends1941} 
demonstrates clearly that behavior is underlain by a network of modules. A normal behavioral sequence of the wasps is as follows: Build a nest; close the entrance temporarily with soil; fly away and hunt a caterpillar; carry the paralyzed caterpillar back; reopen the nest; put the caterpillar in; lay an egg; close the entrance again, this time with greater care. Now build another nest and repeat the entire process so far. Now return to the first nest; open the closure; make an inspection visit. If the egg has hatched and the nest is in order, close the entrance, and now bring 1-3 caterpillars in succession. Repeat this second phase for the second nest. Now return to the first nest, open the closure and make an inspection visit. If all is in order, bring 3-7 caterpillars in succession; then make an especially careful final closure of the nest entrance. Repeat this third phase for the second nest. Now build another nest, and repeat all from the beginning. Furthermore, if, in the first inspection described above, the egg has not hatched, the wasp may build another nest at that time. It can manage 4 nests at a time, with offspring at different ages at each nest requiring different amounts of provisioning (based on information obtained in inspection visits). If a nest has been disturbed, the wasp may abandon it. 

\begin{figure}[pdf] \centering{
\includegraphics[trim=0cm 4.7cm 0cm 0.05cm, clip=true, width=1\textwidth]{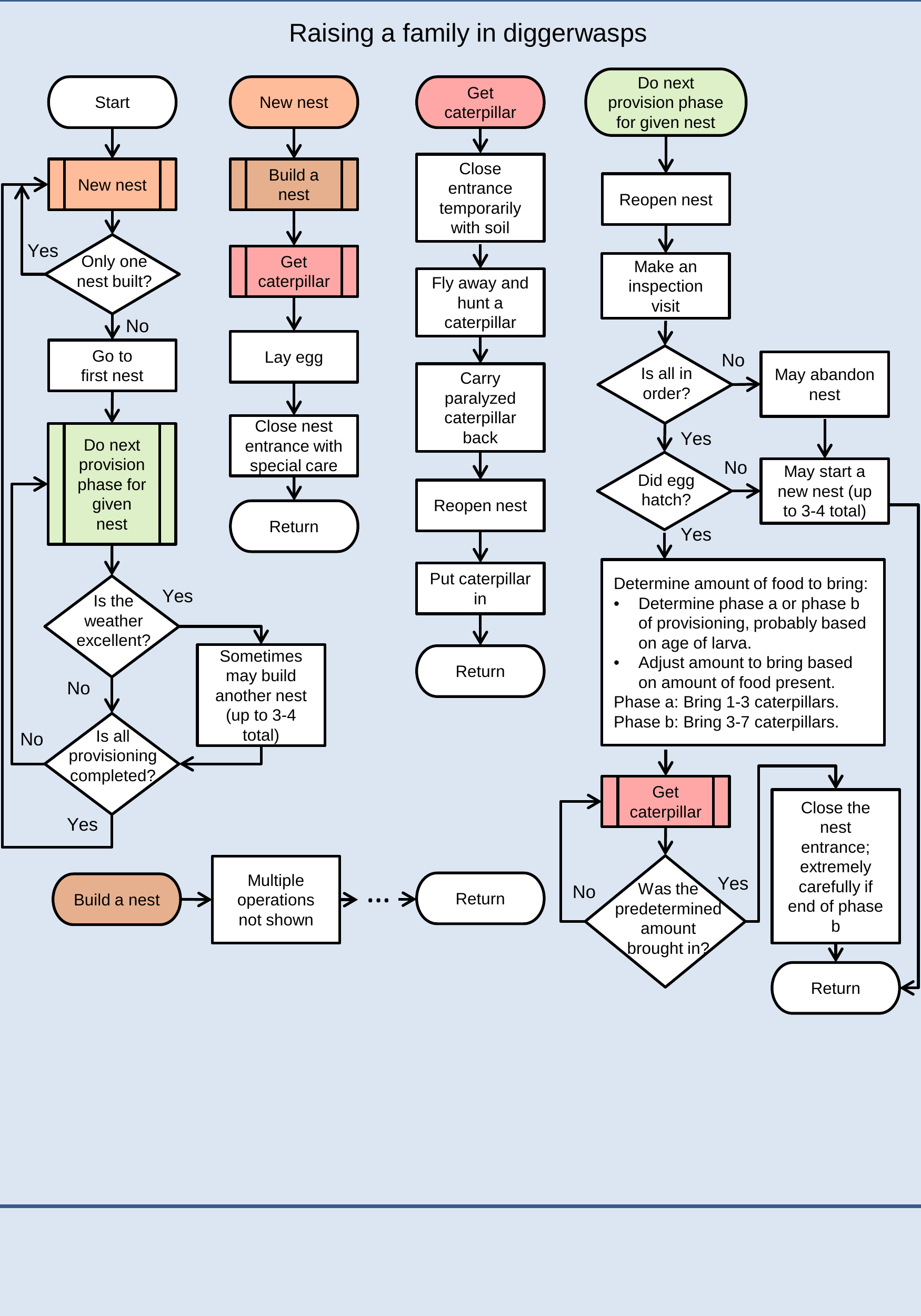}
}
\vspace{-3mm}
\caption{A flowchart describing the algorithmic behavior of nest building, egg laying and provisioning in the digger wasp, \textit{Ammophila adriaansei}. The main procedure begins with ``start;'' next to it appear the subroutines ``New nest,'' ``Get caterpillar'' and ``Do next provisioning phase...;'' and subroutine calls are denoted by rectangles with double vertical edges. Importantly, the entire apparatus is innate and must have evolved somehow (learning is very limited, and is involved in the acquisition of local orientation but not in the behaviors in the flowchart \cite{Lorenz1981}). Interspersed with the activities in the flowchart, the wasp may forage for herself or sleep (not shown). Note that the flowchart is only an approximation, albeit a close one, because of minor incomplete details in Baerends's \cite{Baerends1941} description (such as whether the wasp builds two nests or one after finishing provisioning all existing nests, and several other details).} 
\label{waspflowchart}
\end{figure}

A computer programmer would instantly recognize that the digger wasp's behavior is \textit{an algorithm with subroutines} (see flow chart in Figure \ref{waspflowchart}). The most parsimonious description of this behavior involves activation of the same behavioral modules or subroutines, such as ``carry a caterpillar to nest'' or ``perform an inspection visit'' in different contexts, and the different contexts, namely the different stages of laying and provisioning, themselves consist of different combinations of lower-level behavioral modules \cite{Lorenz1981}.  

Interestingly, Tinbergen wrote that the process underlying emancipation was not known, though it must somehow involve natural selection \cite{Tinbergen1952}. The present theory highlights how correct he was to emphasize that unknown. At once we can understand the inability of traditional evolutionary theory to explain empirical observations from ethology: A network is defined by interactions. The evolution of a network is the evolution of a complex whole. The conceptualization of evolution based on traditional theory encouraged a one-trait-at-a-time type of thinking and was not suitable for discussing network evolution and the transfer of an element from one context to another. Importantly, Tinbergen also noted that there is no point during emancipation at which a behavioral element stops belonging to its original function and starts belonging to a new function \cite{Tinbergen1952}. Rather, as in the evolution of language and in the verbal model of network evolution discussed earlier, the change of context and meaning is gradual. 

Let us now think about the evolution of a network such as described by Baerends. Obviously, elements were not added to it in the form in which they exist today. For example, the construction of a well-shaped nest with a cell at the end had been preceded by a less involved modification of the environment. Also, elements were not appended in the course of evolution at the end of the behavioral sequence. That is, if ``build nest'', ``make closure,'' ``hunt caterpillar,'' etc., are denoted \textit{a}, \textit{b}, \textit{c}, etc., then it is patently obvious that the stages of evolution did not proceed in the following sequence: \textit{a}, \textit{ab}, \textit{abc}, etc., or else absurdities arise such as not laying eggs until a certain point in evolution, performing an inspection visit before the existence of a foraging stage where information from this visit is used, etc. This means that the behavioral sequence was reorganized in the course of evolution and/or new elements were added at internal spots in the sequence. It follows that elements that came after spots into which new or preexisting elements were inserted, or from which preexisting elements were removed or translocated, must have been emancipated from their previous triggers (namely the completion of the behavioral steps that used to come before them) and subjected to new triggers (the completion of the behavioral steps that come before them now). Finally, we would not assume that each repeating element or subroutine evolved afresh in its entirety for each instance in the sequence in which it is used. This means that there has been a copying of routine calls, or, to use more generic terms, \textit{copying and differentiation of modules}. Thus, operators of network evolution---emancipation, cooption, copying and differentiation of modules---have been involved in the evolution of digger wasp behavior.  

Another example showing the insertion of elements at internal points in a sequence and sequence reordering is Lorenz's study of display sequences in surface feeding ducks \cite{Lorenz1958}. Lorenz found about 20 behavioral elements, of which different combinations make different display sequences in different species and even within the same species. Lorenz \cite{Lorenz1958} describes the study of three different species, the mallard, the European teal (\textit{Anas crecca}) and the gadwall (\textit{Anas strepera}), which share the following 10 elements: 

\begin{enumerate}
\item Initial bill-shake
\item Head-flick
\item Tail-shake
\item Grunt-whistle
\item Head-up-tail-up
\item Turn toward the female
\item Nod-swimming
\item Turning the back of the head
\item Bridling
\item Down-up movement
\end{enumerate}

He then presents some display sequences (where one element follows another in quick succession) for each of the three species. For the mallard: 
\begin{itemize}
\item 3,2,3
\item 1,4,3
\item 5,6,7,8
\end{itemize}

For the European teal:
\begin{itemize}
\item 3,2,3
\item 10
\item 4,3,2,5,6,8
\end{itemize}

For the gadwall:
\begin{itemize}
\item 4,3,2,3
\item 5,6;10,6
\end{itemize}

(the semicolon mark between sequences 5,6 and 10,6 in the gadwall means that they are welded at high excitation, which suggests, by connection with many other observations, that we are observing them in the midst of a process of evolutionary welding \cite{Lorenz1958}.) 

The sequences above are obligatory and innate. Hybrids produce their own sequences. This clearly shows network-level evolution in the sense of reorganization of modules, emancipation and welding at the level of sequences of fixed action patterns (FAPs). 

Critically, this example and the previous ones show us that \textit{the picture that we obtain by looking closely at evolution at the phenotypic level mirrors what the molecular biological and genomic revolutions have taught us about the genetic level:} both at the molecular and at the phenotypic levels, network-level evolution is key. And network-level evolution is much better understood with the help of the principles of interaction-based evolution, including cooption, emancipation, acceleration and simplification.  

Welding, like emancipation, is also a network-evolution operation. While Baerends's and Lorenz's examples above demonstrate it at the level of sequences of FAPs, welding can also generate elements at a lower level, namely the FAP itself; though---critically---and in accord with our earlier discussion of network evolution at the molecular level---there is no sharp boundary between the FAP and sequences thereof. One telling example was the inciting ceremony in ducks \cite{Lorenz1971,Lorenz1958,Lorenz1966} described in section \ref{incitingceremony}: To-and-fro movements of the female duck, originally triggered by separate environmental stimuli, have gradually fused in evolution and have become triggered as one, while becoming emancipated from the presence of neighbors. These movements originally were a territorial behavior, with an indirect, implied meaning of pair-bonding and team work, and as they fused, the pair-bonding meaning crystallized and moved to the fore \cite{Lorenz1971,Lorenz1958,Lorenz1966}.    
In fact, many related examples exist; for instance, the territorial marking in the fire-mouth cyclid, \textit{Cichlasoma meeki}. The tendency to attack a neighbor when in one's own territory and flee from the neighbor when in the neighbor's territory is indeed a very general one, spanning birds, fish and mammals. In some fish, the neighbors exchange chase and be-chased turns, coincident with crossing the territorial boundary \cite{Lorenz1981}. In the fire-mouth cyclid, this chase and be-chased movement has become a highly rhythmic oscillation---it has become stereotyped. The welding of the previously separately triggered back-and-forth movements in this species is revealed when one fish suddenly loses interest and disengages yet the other continues oscillating \cite{Lorenz1981}.   

It is due to the highly surprising nature of these examples that Lorenz has been accused of Lamarckian thinking, even though he rejected it. The problem is that traditional evolutionary theory is not network based, and thus it has been impossible to properly conceptualize these examples from its perspective.

\subsection{Generalizing beyond ritualization: the automatic nature of instinct} \label{automaticnatureofinstinct}

The following examples not only show that welding and other elements of network evolution extend beyond ritualization but also demonstrate the automatic nature of instinct. Consider, for example, the pecking instinct in domestic chicks. This FAP is present at birth and consists of three main elements: lunging the head, opening and closing the beak, and swallowing \cite{Lehrman1953c}. Since we would not assume that these three elements of the fixed action pattern have each evolved from scratch in the context of this FAP, we are forced to assume that they have been welded. 

The classic example of a FAP---egg rolling in the greylag goose (\textit{Anser anser})---also shows welding. Upon seeing an egg placed by the side of its nest, the goose stretches its neck in a particular fashion, places its beak over the egg, and then slowly rolls the egg back into the nest while performing balancing sideways motions with the beak to prevent the egg from slipping from the side \cite{LorenzTinbergen1970}. This seems like an insightful sequence of operations, but in fact, when the egg is quickly pulled out from under the beak while in motion, the goose will continue to roll the remaining nothingness all the way to completion and tuck it under \cite{LorenzTinbergen1970}, again showing the automatic nature of instinct. 

Indeed, it is implicit in Barlow's definition of the FAP (fixed action pattern) that the FAP is a welding of elements in general \cite{Barlow1977}. Barlow's definition of the FAP, which he renamed ``modal action pattern'' (MAP), is that it consists of a behavioral module \textit{usually indivisible} but \textit{made of elements that appear individually elsewhere.} Relatedly, Lorenz had suggested \cite{Lorenz1958} that \textit{``perhaps all behavioral patterns'' arise from welding} such as seen in the inciting example. 

\subsection{Ritualization shows all elements of network evolution in one}

Interestingly, a single case of ritualization often exemplifies multiple or even all of the following characteristics: emancipation (also: routinization, autonomization, or evolution of innateness), cooption, chunking (or welding), increased efficiency, exaggeration (or caricaturization), schematization, simplification, stereotypy, automatization and rigidification \cite{Huxley1914,Whitman1919,Armstrong1950,Tinbergen1952}. Notably, traditional theory has only offered to explain one or another of these phenomena in separate from the others. For example, Maynard-Smith and Harper \cite{Maynard-SmithHarper2004} suggested that stereotypy evolved because it standardizes competition, which not only ignores the co-occurrence of the many elements above-mentioned, but also ignores the fact that stereotypy exists also in non-signaling instincts. In contrast, it is striking that interaction-based evolution unifies all of these observations under one umbrella, as outlined below: 

\begin{itemize}
\item \textbf{Emancipation.} Emancipation (the release of a module from previous influences), or the evolution of innateness, is clearly demonstrated by the examples above, and has been addressed here as a part of interaction-based (or network-based) evolution. The same is true for chunking---the combining or welding of modules---which is also a part of network-based evolution.   
\item \textbf{Simplification.} A fundamental concept in ethology is that of the ``sign stimulus''---the stimulus that elicits a fixed action pattern. Here, ``sign'' means ``simple'': the sign stimulus obtained its name from the fact that the animal attends only to a very limited part of the situation that we know it to be capable of perceiving through its senses. Namely, it attends to a parsimonious summary of the situation---a key. Yet the simplicity of the key is only relative: it is still a complex whole, a pattern involving relations between elements 
\cite{TinbergenKuenen1939,Lorenz1939,Kratzig1940,Tinbergen1951}. 
For example, the gaping response of nestling \textit{Turdus merula} as soon as they open their eyes can be directed at a model consisting of a mere three discs that touch each other. However, it is preferentially directed toward one of the discs that bears the right size-relation to another disc, such that the two together can be interpreted as head and body \cite{TinbergenKuenen1939}. As another example, an abstract cross-like model (including symmetrical anterior and posterior ``wing'' edges, and central short and long protrusions perpendicular to them) elicited an escape response from young birds, but only when it is moved in the direction of the short end of the cross, as only in this case the short end can be interpreted as a short neck, which is the case for birds of prey \cite{Lorenz1939,Kratzig1940,Tinbergen1951}. In other words, \textit{an abstract combination of elements} is the evolved key. Now, Tinbergen argued that evolved rituals, which are themselves stimuli eliciting behavior in others, have been ``schematized'' through evolution and are evolved sign stimuli \cite{Tinbergen1952}. Thus, he implied that rituals (and I will add the reception of signals, the ``innate releasing mechanism,'' or IRM \cite{Lorenz1981}) have been evolutionarily simplified to their complex essence. \item \textbf{Exaggeration.} Ritualized signals are often exaggerated, as in the case of throwing the neck over one shoulder and then the other during inciting in the golden-eye (section \ref{incitingceremony}). 
Although it has been suggested that exaggeration has evolved under natural selection for visual clarity, it is questionable that organisms would need such a degree of clarity\footnote{As an example of the animals' acute discriminatory abilities, a herring gull can recognize its mate among a group of other gulls from 30 yards away \cite{Tinbergen1951}}. I argue that exaggeration is related to ``caricaturization'' or ``schematization'' (terms used in the literature) and the final touch of perfection (section \ref{finaltouch}), and evolves by simplification under performance pressure (section \ref{simplification}).   
\item \textbf{Stereotypy.} Stereotypy---or the lack of variation between individuals in a certain trait, or even between different instances of the behavior in the same individual---is another prominent aspect of rituals. According to interaction-based evolution, stereotypy is an inherent aspect of evolution, as will be discussed in section \ref{degreesoffixedness}. 
\item \textbf{Cooption.} Cooption is inherent to Tinbergen's definition of ritualization, as noted (a non-signaling behavior is coopted as a signal) and is also a crucial part of network-level evolution. 
\end{itemize}
Thus, interaction-based evolution provides a much more parsimonious view of ritualization than traditional theory, which provides both additional support for the present theory and an improved conceptual understanding of ritualization.

\subsection{Network evolution, language evolution and phenotypes} 

In the model of network evolution (section \ref{verbalmodel}), I argued that two genetic elements that previously were regulated by two separate lines of controls and had to be separately expressed before coming together in an interaction can, over evolutionary time, gradually come together under one control and even fuse to form a new gene. In this process, there is not only emancipation (one or both of these elements is emancipated from what previously controlled it) but also a sense of automatization, innateness and acceleration, as the emerging unit is no longer constructed from its elements by developmental interactions but rather has been evolutionarily accelerated into a ready-made unit or gene. 

We can now see that the phenotypic-level examples from the previous sections that demonstrate emancipation and cooption also demonstrate the evolution of innateness, automatization, and acceleration, as expected. In the pecking instinct of domestic chicks, for instance, the lunging of the head, the opening and closing of the beak, and the swallowing, have been welded together. Therefore, the last two elements follow the first one now automatically, even though they must have been originally triggered separately by the environment. Furthermore, this welded instinct appears soon after hatching, and perhaps to some degree in the embryo \cite{Lehrman1953c,Kuo1932a,Kuo1932d}, demonstrating the evolution of innateness and acceleration. In the case of egg rolling in the greylag goose, the initial stimulus from the egg suffices to trigger the entire motion of the beak that performs rolling all the way back to the nest even if the egg is removed during its journey. And in the inciting ceremony, the evolutionary process has emancipated the to and fro movements from environmental triggers, welded them together and put them under the control of one trigger, resulting in the evolution of innateness, and has brought the pair-bonding meaning to the fore (section \ref{incitingceremony}). 

Finally, in the evolution of language, we saw that a pair of words can gradually acquire new meaning from the gradual change of the context of its usage, and at the same time can begin to be perceived and learned directly as a new word (word fusion) or concept in and of itself; whereas previously the emerging meaning of it had to be constructed from other words that had to be learned earlier. In this process, there is not only emancipation of the word fusion from its previous context but also a sense of acceleration of the learning of the meaning of the word fusion. This acceleration amounts to automatization of the new concept, which is no longer constructed from other, more elementary units. Thus, the concepts of network-based evolution and the absorption of meaning from context are central to genetic evolution, phenotypic evolution and more. 

\subsection{Stereotypy and the evolution of complex phenotypes: the process by which phenotypes become fixed} 
\label{degreesoffixedness}

Interaction-based evolution holds that selection continually operates on complex interactions between alleles across loci \cite{Livnat2013}. Thus, mutations do not generally bring independent pieces of the phenotype from the individuals in which they originated to all, as they are required to under traditional natural selection. Instead, they interact, and the phenotype evolves at the level of the population as a whole \cite{Livnat2013,LivnatEtal2014a}. It follows that, as some genetic variation across loci gradually disappears (even while new genetic variation appears elsewhere), the phenotypic variation that it caused due to the sexual shuffling of the genes disappears, and thus, over the generations, parents and offspring gradually become more similar to each other. In other words, interaction-based evolution necessitates that a trait gradually becomes stabilized at the level of the population as a whole---it becomes ``fixed'' at the phenotypic level. This concept, called ``convergence'' in \cite{Livnat2013}, 
shows us that if evolution is based on interactions between alleles across loci, then it must involve stabilization and therefore stereotypy as an inherent part of the process. 

If traits are gradually stabilized in such manner, then it follows that we should see a continuum of phenotypic fixedness corresponding to the formation of traits, with different traits lying at present at different points along that continuum. Some traits, still early in the process of formation, will appear as less stereotyped, and others, at later stages in the process, will appear as more so. Thus, through this lens, \textit{stereotypy can be viewed as an indication of the degree of evolutionary progress of a trait.} 

An example is provided again by the pointers. 
Darwin wrote that the hunting behaviors that characterize the pointers are basically innate, and that the only difference between them and ``true instinct'' is that they are ``less strictly inherited'' in that there is variation in the individuals' ``degree of… inborn perfection'' and therefore in the extent to which they require training \cite[p.237]{Romanes1883}. Indeed, pointing in a statuesque manner, backing other dogs
and other hunting-relevant behaviors have all been observed to occur often in pups that have not had the opportunity for learning by instruction, imitation or experience, and they are not exhibited immediately or to the same degree in all pups \cite{Arkwright1902}. When training is required, the amount of training required is small: the trainer only guides the dog toward expressing what it has already a strong natural tendency to express \cite{Arkwright1902}. The existence of variation in pointing behavior becomes eminently natural from the perspective of interaction-based evolution: if traits evolve from fuzzy to sharp, if they gradually become stabilized, as discussed in \cite{Livnat2013}, then this simply means that pointing is still in the process of formation and has not yet become perfectly innate. 

Those who previously tried to explain the fixedness that is stereotypy tended to argue that signals must be clear, that stereotypy makes them clearer, and that they are selected for this extra clarity in a traditional process of selection, hence stereotypy \cite{Barlow1977,Mayr1974}. However, the early ethologists began by studying an extreme---the fixed action pattern---and later, Barlow noted that even what were previously called FAPs are not uniformly uniform, but rather some FAPs vary more than others; they are not all completely ``fixed'' \cite{Barlow1977}. To explain this continuum of stereotypy, it has been proposed that signals that need to be clearer are more stereotyped, and others are less so \cite{Barlow1977}. However, I argue that the clarity-based approach lacked parsimony from the beginning, because stereotypy is a property of instinct in general, not just of signaling behavior specifically; and that even if clarity plays some role in the evolution of stereotypy, uniformity varies first and foremost because of the temporal nature of the process. That is, the degree of stereotypy is in general associated with the point that the trait has reached along the spectrum of formation, and we observe that it varies across traits because we are witnessing traits at different stages of formation. In other words, even if we can imagine reasons why uniformity \textit{per se} would also be of value, the traditional focus on uniformity as a separate end obscures the general point of interest: stabilization is an inherent part of interaction-based evolution\footnote{This is not to say that all traits must inevitably cover the whole spectrum and reach the extreme, nor that they all move along the spectrum at the same rate; but it is to say that the degree of fixedness comes from the nature of the process of network-based evolution, and is not an independent element traditionally selected for, as Maynard-Smith and Harper \cite{Maynard-SmithHarper2004} and Mayr \cite{Mayr1974} have argued. Indeed, whether the evolving trait is a signal or not, the evolutionary process is converging on an adaptation: along with the decrease in variance and disorder, it converges on highly efficient structure and behavior. Clarity, which is part of the effectiveness of the signal, and efficiency in other adaptations, are outcomes of the evolutionary process, and uniformity is an inherent concomitant of the process in both cases.}. 

Hinde's comparative study of displays in finches demonstrates several of the points above \cite{Hinde1955}. First, the same trait may vary more or be more stereotyped in one species than another. Second, demonstrating the evolution of FAPs from interactions, most displays in the finches studied by Hinde are not yet fixed action patterns but rather are poses whose elements tend to occur together statistically. Third, of the different displays, the one which is particularly rigid, stereotyped and emancipated (the female soliciting posture), is also the one that is far more widely shared, and therefore more ancient, than the other, still variable displays,
in accord with the prediction that stereotypy is indicative of the evolutionary stage of formation of a trait. Another, in-depth example showing the evolution of different degrees of fixedness, and that stereotypy is a concomitant of the evolution of adaptation, namely the evolution of decoy-nest construction in sand wasps, will be given in section \ref{Evans}. 

\subsection{Interaction-based evolution provides a single explanation for the different aspects of innateness}
\label{unitesdifferentaspects}

``Innate behavior'' or ``instinct'' has been used to mean different things in the literature \cite{Papaj1993}:
\begin{enumerate}
\item Independence: It has been used to refer to behavior that is \textit{independent of interactions with the external environment} like learning or experience. Such independence is especially clear when a behavior is present from birth, as for example in the pecking of domestic chicks \cite{Lehrman1953c}\footnote{ ``Environment'' and ``experience'' need to be qualified: No behavior is entirely independent from the ``environment'' or from ``experience'' when these notions are so broadly construed as to include such factors as the flow of food materials during development or the ``source of experience'' that one body tissue provides a neighboring one in the course of development. Even what we more normally call ``experience,'' that is interaction with the outside during growth and learning, can figure into and change aspects of behaviors that are otherwise innate. For example, the tendency of the rat to build a nest is innate, yet it will not be able to carry it out if it is deprived of experiencing the carrying and manipulating of objects during development \cite{Lehrman1953c}. Despite this important qualification, the conceptualization of innateness as independence from the environment is still true and useful in an important sense.}. 
\item Stereotypy: Innate behavior has been referred to as \textit{stereotyped} (also, ``fixed,'' ``constant,'' ``rigid,'' \cite{Papaj1993}---``robotic'' ). 
\item Sharedness: Innate behaviors have been found to be homologizable between species and therefore useful for taxonomy. In other words, they are \textit{shared between species}, with some species-specific characteristics \cite{Lorenz1981,Heinroth1911,Whitman1898,Whitman1919}.
\end{enumerate} 

The fact that these three aspects are empirically connected is clear. First, both Tinbergen and Lorenz considered it of great importance that stereotyped behavior (fixed action pattern; point 2 above) is also homologizable between species and therefore is as useful for taxonomy as morphological characters are (point 3) \cite{Tinbergen1951,Lorenz1981}\footnote{One must note, however, that parallel evolution of stereotyped traits in related species could also lead to ``homologizable'' traits in this sense even though in this case there is no common origin, strictly speaking, a scenario which is in fact expected from interaction-based evolution.} (Lorenz called it ``an epoch making discovery'' \cite[p.103]{Lorenz1981}).  
Second, there is an agreement that, empirically, what is innate in the first sense above (1)---what is automatic---appears also to be stereotyped, or ``fixed;'' it appears ``robotic'' (2)\footnote{Even though there is variation in the degree of stereotypy, there is also variation in the degree of independence from the environment, and biologists agree that what is strongly fixed tends to be strongly uninfluenced by the environment, or ``innate'' in the first sense \cite{Papaj1993}.} \cite{Barlow1977,Papaj1993}. 

But why are the different aspects of innateness connected? Tinbergen's tone regarding the first connection outlined above (between points 2 and 3) was that of surprise \cite[p.191]{Tinbergen1951}. And Barlow discussed stereotypy in the context of the clarity of signals \cite{Barlow1977}, which neither addressed stereotypy in non-signaling behavior nor explained the connection between stereotypy and independence (even though his continuum of FAPs ran both from variable to stereotyped and from dependent on to independent of the environment at the same time \cite{Barlow1977}). 
Finally, Papaj's model was unable to connect stereotypy and independence \cite{Papaj1993}. 

However, from the point of view of interaction-based evolution, the different aspects of innateness are connected. First, traits evolve by a process of convergence as defined (see section \ref{degreesoffixedness} and \cite{Livnat2013})---a process of stabilization  
which begins at a state of high variance and eventually leads to the evolution of uniformity and therefore stereotypy \cite{Livnat2013}. Since this process takes time, it makes it so 
that stereotyped elements are older and more widely shared than elements not yet stereotyped, while implying that stereotypy evolves in parallel. \textit{This point connects the sharing of a trait among species with stereotypy} (aspects 3 and 2 above, respectively).

Second, I have argued that interaction-based evolution works in the long-term through simplification under performance pressure. This can not only make one intra-organismal module independent of another, but also make it independent of environmental factors (see section \ref{modularityandinnateness}). Therefore, besides leading to stereotypy, the process of interaction-based evolution also leads to increased  independence from environmental factors (1). 
\textit{This ties together independence, phylogenetic sharedness and stereotypy.}  

Two additional aspects of innateness also make sense in light of interaction-based evolution. One is that welded elements often and perhaps always \cite{Lorenz1958} are present in the fixed action pattern (see sections \ref{elementsofnetworkevolution} and  \ref{automaticnatureofinstinct}), which aligns with the fact that welding is an outcome of network-level evolution. It also sheds light on the issue of circuitous vs. accelerated development, as will be discussed below (see section \ref{howevolutionlearns}), which will clarify the connection between evolution and learning. 

Traditional evolutionary theory has had difficulty explaining and reconciling the various aspects of innateness. This has resulted in a call by Bateson to give up the use of the term ``innate'' altogether and specify instead what particular meaning of innateness one is referring to \cite{Bateson1991}. In contrast, by connecting the different aspects of innateness with recourse to one parsimonious mechanism, interaction-based evolution fits with the fact that the term ``innate'' has been used intuitively as a unifying concept for a long time. Furthermore, it shows \textit{how} these different aspects are related to each other.

\subsection{Evolution from fuzzy to sharp: examples of the evolution of complex phenotypes}
\label{fuzzytosharp}

Because of their \textit{ostensible} simplicity, examples 
like the evolution of malaria resistance due to the HbS mutation have served the traditional notion of natural selection and random mutation. In fact, fascinating molecular-level details are available that question the accidental nature of this mutation and others \cite{Livnat2013}. However, putting aside this mounting evidence, another fundamental issue is that examples of 
the evolution of complex phenotypes have been 
very much underplayed in the evolutionary theory literature. In this paper, we have seen that these examples fit with the view of interaction-based evolution: they demonstrate cooption and emancipation, stabilization and stereotypy, and evolution from fuzzy to sharp. Most intriguingly, they \textit{speak to the arising of novelty in a way that is consistent with network-level evolution}. 

For the reader who is interested in the detail, I will first discuss the example of decoy construction in sand wasps. Other readers may skip to the final and most important phenotypic-level example---that of egg retrieval by backward walking in the nightjar (section \ref{backwardwalking})---where I will demonstrate all the elements discussed in this paper in one, with an emphasis on the central point of novelty. 

\subsubsection{Elements of network evolution in the construction of decoys in sand wasps} 
\label{Evans}

This example is taken from Evans's classic work on the comparative ethology of the sand wasps, Bembicinae (previously called Nyssoninae; \cite{Evans1966a}). To make a nest, the sand wasps dig a tunnel of many body lengths, at the end of which they build a cell or a complex of cells, where they place their offspring and the prey on which the offspring feed. They have natural enemies---two taxonomic groups of flies (the bee flies, Bombyliidae, and miltogrammine flies, Sarcophagidae) and two taxonomic groups of parasitic wasps (the cuckoo wasps, Chrysididae, and ``velvet ants,’‘ Mutillidae)--- that parasitize their nests by laying there, and whose larvae either takes up valuable resources or destroy the larvae of the sand wasp. Parasites of the former two groups seek the nests of their hosts by sight, and of the latter two groups by touch and odor---they fly over the ground, tapping the soil with their antennae in search of their target. Across the sand wasps we find various techniques of hiding and concealment as well as a decoy construction technique. 

The example concerns the decoy construction \cite{Evans1966a,Evans1966b}): Some species of sand wasp dig a false burrow (or multiple such burrows) next to the real one and leave it (or them) open, while leaving the real nest burrow closed. Various parasites have been observed to either lay their eggs or linger in the decoys. Evans hypothesized that false burrows originated in a behavior that had a different purpose---that its origin is in cooption---and that the process of the evolution of digging false burrows was a process of improvement through stereotypy, together with emancipation (see also \cite{Tsuneki1963}). 

He based his hypothesis on the following facts. Very commonly across the Bembicinae, the wasps close the entrance to their nest from the outside before leaving it, either temporarily with a small amount of soil before leaving temporarily for provisioning, or with a large amount of soil at the final closure before leaving for good; and both within and outside of the Bembicinae, species have been observed where individuals obtain soil for nest closure mostly from several or one particular spot/s around the entrance. In this case they leave behind a small pit or pits of a size that depends on how much a particular spot was used. The tendency to take soil from a particular spot or spots appears to relate to environmental conditions, where individuals quarry soil for closure when loose soil is not as easily available \cite{Evans1966b}, though it also has a genetic component \cite{Evans1966b}. The more soil was quarried from a particular spot, the bigger the pit left behind. In \textit{Bembecinus neglectus}, for example, most individuals took the soil for closure from several particular spots around the entrance, so that a ring of small depressions was left behind. But some took it mostly from one particular spot, which formed a ``depression or short ‘false burrow’ up to 1 cm deep'' \cite[p.137]{Evans1966a}.

Now, in the genus \textit{Bembix}, which typifies the advanced behaviors of sand wasps, we see species with decided false burrows. Three species are particularly telling: \textit{Bembix amoena}, \textit{B. texana} and \textit{B. sayi}. In \textit{B. amoena}, false burrows are of irregular occurrence and spatial pattern and are relatively short, and the burrowing is still almost always associated with quarrying soil for closure. Most individuals obtain soil for the initial closure by scraping a small amount of soil from each side of the nest entrance, and occasionally from a particular spot or spots, creating short false burrows that stretch in a direction of 90 degrees or a bit less left or right of the direction of the true nest, obliquely into the ground. These short false side burrows varied in length from barely perceptible to 2 cm long, with two exceptional cases observed at 3 cm and 5 cm long \cite{Evans1966a}. Evans \cite{Evans1966a} noted that, in one population, about half of the nests had no such false side burrows, about a quarter had one, and the rest had two such false side burrows at one time or another. In addition, individuals also collected soil from opposite of the nest entrance, most often, but not always, for the final closure (which requires more soil), which resulted either in a furrow going through the mound of soil opposite the true nest (a mound resulting from the excavation of the true nest), or in a burrow running under that mound obliquely through the ground. The former appeared a considerable number of times, \cite[ p.280]{Evans1966a} and were of varying length, between 1 and 7 cm long. The latter appeared rarely \cite{Evans1966a}, with only three cases noted, at 1.5, 2 and 3 cm long. Evans \cite[p.281]{Evans1966a} noted that two of these were made \textit{following} final closure in a manner similar to that of \textit{Bembix sayi} (see below), which seems to suggest that these two rare instances occurred not in the service of obtaining soil for closure. 

Besides the burrows themselves, the manner and timing of construction of the burrows was also variable of course (Evans 1966a,b). The burrows on the sides, when they occurred, occurred sometimes along with the initial closure and sometimes along with later closures. The back burrows and furrows, when they appeared, appeared often along with the final closure but sometimes along with earlier closures, and the final closure did not always involve them. While burrows were sometimes revisited and expanded, they were sometimes accidentally filled while making a closure. Likewise, the spatial pattern resulting at the end was variable, with 0 or 1 or 2 side burrows and 0 or 1 back burrow or furrow \cite{Evans1966a,Evans1966b}. Note that, although the soil was generally taken for closures, parasites were distracted by the false burrows that resulted \cite{Evans1966b}. 

In \textit{B. texana}, construction of false burrows is more or less regular, and the soil is not used for closure. Typically, individuals construct one short but relatively persistent false burrow on each side of the entrance, right after the initial closure is made \cite{Evans1966a}. The method of construction is not yet entirely stereotyped, with some individuals digging one burrow and then the other, and others alternating in digging both \cite[p.325]{Evans1966a}.

In \textit{B. sayi}, construction is invariable and emancipated: all females dig one strong back burrow 4--22 cm long under the mound after completion of final closure (which also means that the soil is not used for closure) \cite{Evans1966a,Evans1966b}.    

The three species above exemplify certain general trends \cite{Evans1966a,Evans1966b}. The primitive cases of false burrows, where the burrow is but a small pit, are unreliable and irregular in appearance. The ``transitional'' case of \textit{B. amoena} shows burrows appearing in a notable number of cases but still rather irregularly. They are often longer than ``small pits'' but are still relatively short and vary greatly in length. In the advanced cases, the burrows appear with greater regularity and are substantial. In \textit{B. sayi}, which makes the longest burrows, the burrows appear invariably in all individuals and at a regular time. There is an association between stereotypy and completeness of the burrow \cite{Evans1966a,Evans1966b}. 

In addition, there is an association between stereotypy and emancipation of burrow-making from its previous cause \cite{Evans1966a,Evans1966b}. That is, limited quarrying in the service of obtaining soil for closure is a widespread phenomenon, and tends to be irregular in occurrence; whereas, in contrast, regular burrows are the result of burrowing for its own sake, an operation not used for closure, and are constructed at regular times before or after closure, depending on the particular species concerned. In some of the advanced species where burrow-making is thus emancipated, the wasps refresh or fix burrows that have been destroyed \cite{Tsuneki1963}, which further clarifies that they are programmed to maintain a certain pattern of false burrows. The above characteristics are also associated with increased regularity of the spatial pattern of the burrows, with each of the different emancipated species having its own idiosyncratic characteristics of construction \cite{Evans1966a,Evans1966b}. 

Furthermore, not only have emancipated burrows never been observed in species that lack closures, but in addition, a careful reading of Evans \cite{Evans1966a} shows that, even though they are no longer used for closure, emancipated back burrows are temporally associated with final closures, and emancipated side burrows are temporally associated with initial closures, which seems to cross-validate the fact that the origin of emancipated burrows is in closure-making. 

Thus, evidence clearly supports the predictions of interaction-based evolution. The evolution of false burrows \textit{originated in cooption---in emerging high-level interactions between preevolved elements like digging, quarrying and making closures, and environmental elements like sand conditions}. That was a state of high variance in behavior and outcome within and between individuals. Evolution then proceeded \textit{from fuzzy to sharp:} through a process of convergence and gradual stabilization of the trait as a whole \textit{toward a stable, emancipated and clock-work--like state}. The process was that of improvement together with and at the same time as stereotypy and emancipation. Note that it is not the case that complete but irregular burrows evolved first, and then were stabilized. That is, stereotypy, or uniformity, is not an outcome of a force of stabilizing selection separate from the selection for the adaptation itself. Rather, \textit{stabilization and improvement evolve together as two aspects of the same coin---as inherent concomitants of the adaptive evolution of the whole as a whole}, as predicted by interaction-based evolution.  
   
\subsubsection{The emergence of novelty in the evolution of egg retrieval by backward walking}
\label{backwardwalking}

I will now discuss the final and most important phenotypic-level example that puts all of the elements of the theory discussed here together, while emphasizing the central point of the emergence of novelty. This is the example of the evolution of egg retrieval by backward walking in the nightjar (\textit{Caprimulgus europaeus}) and other species, which applies to eggs that have rolled far outside the nest. Before we can understand it, I must first explain what the shifting motion in birds is. 

The shifting motion in birds is ancient and involves rolling an egg with the beak until it reaches under the body. The egg may have thus gotten in between other eggs and stirred them, and the egg sides that are pointing up are thus changed \cite{Tinbergen1960}. Shifting may be needed to ensure even temperature distribution to the eggs \cite{CaldwellCornwell1975}, and is performed upon arrival at the nest, or when the tactile stimulus provided by the eggs while brooding is not satisfying, or spontaneously after a long spell of quiet brooding \cite{Tinbergen1960}. In terns, the shifting motion will move an egg about 2--3 inches. 

Coming back to our case of egg retrieval, in terns (e.g., \textit{Onychoprion fuscatus}), the general situation is as follows \cite{WatsonLashley1915,Tinbergen1960}: If they notice an egg lying several inches outside of the nest, they leave the nest right away to it. However, they have an aversion toward being far from the nest, induced by their brooding state, and as they move away from the nest, they slow down, sometimes turning around and returning to the nest without having reached the external egg. But sometimes they do get to the egg, stopping short of it just close enough that they can reach it with the beak and apply the shifting motion to it, which rolls the egg until it is under the breast. 

As they shift the egg, they sit down on it to incubate it, but only for a short time (indeed they may at this point be dissatisfied with the tactile stimulus and/or with being outside of the nest). The moment they notice the nest again they stand up and walk to it. In the process, the egg has moved about 2-3 inches toward the nest due to the shifting motion. 

Having returned to the nest and started brooding the eggs there, they soon notice the external egg again, venture out toward it again, and repeat the process, and the egg moves 2-3 inches again toward the nest. Thus, after several trips, the egg finds its way back to the nest. 

The behavior that results in the egg being moved back to the nest is clearly unstructured. The brooding of the external egg outside of the nest and the back-and-forth trips show the lack of insight or ``analysis of the situation as a whole'' \cite[p.83]{WatsonLashley1915}, as the different actions taken in the situation are under the proximate control of different preevolved instincts. In accordance with Tinbergen \cite{Tinbergen1960}, these instincts are competing with each other for expression: the desire not to leave the nest, the desire to return to the nest, the desire to brood eggs, and the desire and ability to shift an egg. Also, as Marshall noted for the common tern (\textit{Sterna hirundo}) \cite{Marshall1943}, there is much variance in the behavior and its outcome, with eggs sometimes being rolled back into the nest and sometimes not, and this variance is thought to reflect both individual variance and situational factors \cite{Marshall1943}. 

In other birds, however, such as greylag geese (\textit{Anser anser}), black-headed gulls (\textit{Chroicocephalus ridibundus}) and nightjars, the bird walks straight up to the egg, puts the beak over it as it would in shifting, but instead of incubating the egg there, it then walks backwards all the way to the nest in one shot while shifting and dragging the egg under its beak \cite{Kirkman1937,Tinbergen1960}. 

According to Tinbergen, this egg rolling observed in nightjars and other birds evolved from shifting and other elements of the situation \cite{Tinbergen1960}. Indeed, the fact that the birds are using a shifting motion while walking backward (even though rolling with the wing would have been much more efficient) together with the fact that shifting is ancient, supports this hypothesis \cite{Tinbergen1960}. In fact, Tinbergen notes that the very controversy about whether egg retrieval is an independent adaptation or a by-product of a confluence of instincts in different species shows its route of evolution \cite{Tinbergen1960}.

The argument that this backward walking behavior appearing in nightjars and other birds evolved from a situation akin to that of the terns exemplifies several elements in one: The trait has evolved from fuzzy to sharp; from unstructured and inefficient to structured and efficient; from variable and unstable to stable, stereotyped and ``rigorous.'' In addition, we also see emancipation in it: the return back to the nest originally required the visual stimulus of seeing the nest, but now is triggered automatically as soon as the shifting motion begins and requires no turning-around to the nest. We also see welding: the going-to-the-egg and the coming-back-to-the-nest legs have been welded together in one sequence unleashed by the stimulus of seeing the external egg for the first time, whereas previously they were two separate legs each triggered by its own visual stimulus. The whole situation has been \textit{simplified}, the path has been straightened up. In fact, the simplification has been the creation of a method from a previously non-methodical occurrence, when all the while the whole evolved as a whole, not by the addition of independent elements one at a time. 
 
On top of all of the above, one topic deserves a special emphasis: novelty. The example of egg rolling shows 
clearly that different instincts or elements have the inherent ability to come together into new, useful interactions that together can achieve what had been unachievable before by any one of those instincts alone. Twice we see that this coming together of pre-evolved elements into useful, high-level interactions, which is outside of the purview of the random mutation and natural selection view, breaks a barrier in terms of being able to do something that could not have been done before. First, the confluence of instincts for shifting, brooding and returning to the nest effectively allows the egg to be returned to the nest after several trips, even if in a haphazard way, when none of these instincts by itself is capable of achieving this, nor did any of them originate due to pressure for such egg retrieval. The second barrier broken was this: the invention of backward walking while shifting allows retrieval of the egg in time that is proportional simply to the distance to the egg, whereas the haphazard way only allows retrieval in time that is quadratic in distance. This improvement allows nightjars to retrieve eggs from many yards away, which would not have been effectively possible in the case of terns (indeed terns retrieve eggs from only several inches away). Thus, emancipation and welding have created a behavior that now applies to a broader range of situations than the ancestral traits applied to. Some of these birds now dwell in beaches where eggs can indeed be blown away by wind a great distance. 

These breakings of barriers in the formation of new traits exemplify novelty. The novelty is in the inherent ability of elements to come together into new and useful high-level interactions. These elements come together first in a haphazard state. Their complex interaction then serves as a substrate for simplification under performance pressure, where new such elements will be formed. I propose that this cycle is the heart of the evolutionary process. I have furthermore proposed that it is simplification under performance pressure that is responsible for this inherent usefulness of elements---for their propensity to come together into new, useful interactions that they have not been directly selected for.  

This point provides an understanding of novelty in evolution that is completely different from and not reducible to traditional random mutation and natural selection. The novelty that drives evolution here arises from the coming together of high-level ``modules,'' i.e., from the network as a whole. It is not a local, ``misspelling''-like change at the genetic level. It is not accidental, or random mutation that invents in evolution. Rather, non-accidental mutation and natural selection together process information gradually, and the source of novelty in evolution is the resulting inherent ability of elements to come together in useful high-level interactions, an ability due to simplification under performance pressure.

Watson and Lashley \cite{WatsonLashley1915} saw that the outcome of the haphazard mode of egg retrieval was not intentional. They noted that the egg rolls in the direction of the nest simply because the bird is oriented directly away from the nest as it reaches the egg, so that the shifting motion happens to bring the egg a bit closer to the nest each time. From this they concluded that the egg finds its way to the nest by lucky happenstance. But this lucky happenstance is a far cry from the traditional notion of novelty in random mutation. First, it is not a local accidental mutation that invents, but rather the process starts with high-level interactions, and gradual network evolution creates something new from this source. Second, an extraordinarily deep new question arises. Should we call this source of novelty ``randomness'' or ``lucky happenstance'' at the phenotypic level, and say no more? There is logic to the present situation that goes beyond the purely coincidental. That is, although the egg rolls to the nest only because of the bird's orientation, could the bird have been oriented any other way? Indeed, it is oriented in the way that it is because it is walking straight up to the egg from the nest. Thus, the efficiency of this movement is integrated with the feasibility of the retrieval system. Indeed, the whole situation, while haphazard, is not purely random, but can be seen as a fuzzy sort of ``\textit{shifting in an extended nest}.'' Thus, a confluence of instincts, each useful in and of itself, together give rise to something useful that is different from each of them, but which at first can only appear in a roundabout, highly variable, even though not purely random, fashion. As such, it serves as material for evolutionary simplification and streamlining, which ends up creating something that can be useful in contexts that go beyond the one that originated it. 

It is intriguing that an unqualified notion of the accidental does not sufficiently explain novelty here. When we apply this way of thinking to other cases of cooption, we will see that, individually, they may appear more or less accidental than the above case; but they are not, in general, ``pure coincidences.'' This, together with the question of how exactly simplification under performance pressure leads to inherently useful elements, opens up an intriguing new area for scientific investigation---a science of novelty. 

While others have discussed the possibility of cooption being a source of novelty \cite{GouldVrba1982}, it has 
been discussed within the random mutation view---cooption has been treated as a \textit{random} event and \textit{another} source of novelty \textit{in addition to} random mutation. In contrast, interaction-based evolution argues that cooption is neither a random event nor another source of novelty in addition to random mutation. Rather, \textit{non-accidental mutation and natural selection gradually pave the way to both genetic and phenotypic cooption at the macroscale} through network-level evolution. Thus, 
neither mutation nor cooption are random in the traditional sense even though they produce surprising things, and they are not separate sources of novelty but come together as two inseparable aspects of one process. Cooption is at the heart of the 
process of interaction-based evolution and is built into this process. 

Thus, while in the traditional view, novelty arises by accident at a specific point in space and time, according to interaction-based evolution, novelty is an outcome that arises over time at the network level from the coevolutionary change of many elements. While the drivers of these local changes are not random, these changes still interact with each other globally in a surprising way. Surprise, or novelty, exists, but it is not a mere direct effect of dice rolling. 

It is noteworthy that the tern situation is based on conflict, or competition between tendencies. The bird, on the one hand, acts as though it wants to reach up to the egg and incubate it, but on the other hand as though it wants to remain in the nest. It is also noteworthy that there is individual variation in the overall behavior, and indeed, there may be different ways of increasing the probability of success. One way may be to approach the egg without hesitation. Another may be to get back to the nest without delay once the egg has been shifted. There is an inherent conflict in the situation. Both tendencies have something to contribute, but they are conflicting. To strengthen one at the expense of the other may be harmful. Evolution may need to take a modest though complex step: to find a solution for returning to the nest immediately and only after reaching the egg while engaging it with the beak. This can be achieved, for example, by overcoming the tendency to incubate but only while standing outside of the nest. The relevant rule to evolve, ``incubate in the presence of eggs AND when standing in the nest'' is simple but non-linear. In performing this evolutionary step, the convergence process described in \cite{Livnat2013} may lead to the crystallization of the commonality between successful individuals while resolving the conflict inherent in the situation, making evolution a process of conflict resolution.  

The example also shows us, of course, that the whole is greater than the sum of its parts and that the organism evolves as a whole. Conceptualizing evolution in this way provides an answer to the many inconsistencies that arise from the accidental mutation framework, such as the fact that it often leads us to surmise difficult evolutionary sequences leading to complex adaptations where the intervening steps are not adaptive in and of themselves. 

Finally, the example also shows us a connection between punctualism and gradualism. Suppose that, as the underlying instincts evolve and are being emancipated and adjusted, the balance of tendencies gradually changes such that the tendency to incubate the external egg while outside of the nest falls below the tendency to return to the nest, while the tendency to return remains balanced with shifting, so that returning and shifting are expressed together. In that case, these tendencies may evolve gradually, while the new trait may arise punctually: it may be possible for a bird species to evolve backward-walking retrieval as a whole and rather rapidly, causing a ``phase transition'' at the level of the observed behavior. Backward walking will then appear first in one individual, then in another, then more and more---it will appear ``like the rain.'' Thus, punctualism is better understood when we start thinking in terms of network evolution, as an outcome of gradual trends in the change of a network that interact with each other.  

\subsection{How evolution learns: circuitous vs. accelerated development} 
\label{howevolutionlearns}

The example of the evolution of egg retrieval highlights a fifth aspect of innateness. The terns are not learning to retrieve the egg in the same way that humans learn a task. They have a set of inborn tendencies that, in the situation, result in egg retrieval. At the same time the haphazard behavior which results in egg retrieval does not seem to fit the term ``innate.'' There is another aspect to innateness, and that is the degree to which a behavior develops straightforwardly and quickly in an endogenously driven fashion. Two opposing examples will demonstrate this point. After an emerged butterfly finishes its preparations for flight (like drying its wings), it 
takes to the air and flies in search of food and mates. While the behavior of taking off is instigated by having just completed preparations (so it too is dependent on ``experience'' in some sense), it is 
not driven by their completion, much like a car is not pushed forward by the gas pedal. It is endogenous for all intents and purposes and is a true instinct. In contrast, the behavior which results in egg retrieval in the terns arises circuitously and at a high level, from a meeting of different inborn behavioral elements as well as environmental factors which play a more inherent role in inducing the behavior: the visual stimuli and conflicting instincts do cause the retrieval of the egg. This high level meeting of modules, both internal and external, now serves as the source of evolution from fuzzy to sharp, at the end of which a new innate module, consisting of a combination of the previously independent elements, will arise (that of backward walking). This aspect, namely, how quickly, straightforwardly and endogenously a behavior (or a trait in general, including morphological traits) arises in development is
important for acceleration: In the initial circuity there is a potential for ``straightening up,'' there is a potential for simplification that will lead to the further breaking of barriers (e.g., substantially faster egg retrieval). The environmental factors play a more inherent role in inducing the behavioral outcome in the tern situation than in the butterfly situation (they are less like the gas pedal and more a part of the engine), which means that, in becoming emancipated from them, the life form can ``learn'' from the environment through evolutionary change. That is, it now produces more endogenously what the environment helped to produce before. \textit{I argue that this emancipation is the intergenerational ``learning'' that is done by evolution}, drawing an analogy between evolution and learning (see more in section \ref{evolutionaslearning}).

It is interesting that those who have tried to define innateness often seemed to mean that, in contrast with learned behavior, innate behavior is in a sense ``predetermined'' \cite{Papaj1993}. In other words, in innate traits, the fit with the environment is predetermined, as opposed to learned behavior or morphological plasticity, where the fit is ``acquired.'' However, notice that this predetermined fit \textit{is} adaptation. The evolution of innateness, or automatization, \textit{is} the evolution of adaptation. And, according to interaction-based evolution, the evolution of adaptation involves network-level evolution and the acquisition of a new phenotypic meaning as a result of the changing context in which modules are embedded. Interaction-based evolution shares with neo-Darwinism the reliance on natural selection for evolution of adaptations; it shares with Lamarckism the \textit{appearance} of the inheritance of acquired characters (though it relies on a completely different mechanism); but it shares with neither the new idea that evolution is network-based and interactions-based. 

\subsubsection{The engine of evolution}

Interaction-based evolution argues that the process whereby a population converges \cite{Livnat2013} on an adaptation is a process that converts information from a less orderly to a more orderly state. It proceeds from a fuzzy to a sharp, well-working and stereotyped state.
However, evolution is not \textit{only} a fuzzy-to-sharp process, in that the fuzzy source must first arise. The progress from fuzzy to sharp is therefore only a half of a cycle of the ``engine of creativity'' that is evolution. The other half is that previously made sharp elements come together at a high level to make the new fuzzy source (e.g., the different instincts in the tern situation come together into a disorderly form of egg-retrieval), from which new sharp elements can be made (e.g., backward walking)\footnote{Of course, these cycles do not occur in a sequence one at a time. At any time point in the course of the evolution of a given life form we may expect many co-occurring cycles, each at a different phase.}. I argued that simplification under performance pressure connects the two parts of the cycle. The simple elements it creates not only are improvements but also come together in new complex interactions which serve as the raw material for the next round of simplification. Thus, novelty arises not from accident, but from evolutionary work. 

\section{A new view of the evolutionary process}
\label{newview}

In this section I will revisit the molecular level from the perspective of interaction-based evolution in light of the concepts learned so far. I will clarify the nature of mutation and raise directions for future research regarding it. 

\subsection{Connections between the microscale and the macroscale} 

While \cite{Livnat2013} developed the micro-scale view of network-based evolution, the current paper develops the macroscale view of it. In both, there is a sense of chunking: On the macroscale, genes as well as phenotypes can become welded in the long-term. On the microscale, information from multiple loci comes together in each of many mutational events (including epigenetic changes) in each of many individuals in each of many generations. 

An important remark may now be made. One might think that, according to \cite{Livnat2013}, non-accidental mutation combines information from alleles at multiple loci into one locus in a way that recreates in the mutated locus the combination of alleles as it was. However, this is not what is meant in \cite{Livnat2013}. According to interaction-based evolution, there is a flow of information from the combination of alleles across loci into one locus, which generates a hereditary effect; but this effect does not replicate the combination as is. The situation is analogous to that of a neuron (or a logical gate), whose output does not replicate its inputs yet represents a lasting effect of this combination of inputs that is transferred to the next layer in the network. In other words, interaction-based evolution argues that 
many non-accidental mutational events over many generations and at many loci come together into a network of information flow across the genome and through the generations, from many loci into one and from one locus to many, and this information flow \textit{gradually} leads to phenotypic chunking at the macroscale, among else. That is, the information flow in Figure \ref{flowovergenerations} is the moment-to-moment workings of evolution; cooption and chunking at the macroscale (e.g., gene fusion or phenotypic fusion) are among the long-term consequences of it. 

This point may also help us understand better the gradual manner of occurrence of a gene fusion or of a splicing pattern which, according to interaction-based evolution, are not sudden stochastic events but the results of long-term processes. As discussed earlier, alleles at multiple loci affect the regulation of an alternative splicing pattern, and the information they represent is processed and stabilized in the long term through convergence as defined in \cite{Livnat2013}, thus setting the new alternative splicing patterns and new contiguous sequences that we see today. Many writing events, in many individuals, over many generations, gradually pave the way for network evolution at the gene level (see the example of the fusion of \textit{TRIM5} and \textit{CypA} in section \ref{trimcypfusion}).  

\subsection{The two ecologies working together: the ecology of energy and the ecology of information}

I will now attempt to put the various arguments of \cite{Livnat2013} and of this paper into one philosophical picture. A machine has several aspects: First, it is a finite, unchanging structure that repeats its operation over and over again, performing the same ``trick.'' Second, we tend to think of a machine as something that operates harmoniously and whose parts have been conceived to fit each other harmoniously. Third, novelty or ``out of the box'' thinking is the antithesis of machine-like behavior. 

Now, the traditional idea of natural selection and random mutation is machine-like in the first sense: it is one trick that repeats itself indefinitely without changing its own fundamental nature. That is, random mutation occurs either as an error during replication or for another accidental reason, and natural selection either accepts it or rejects it. The repetition of this operation is traditionally supposed to be responsible for all of life and every innovation in it---a belief that I have argued against. 

Here, I will draw the distinction that the writing of mutation postulated by interaction-based evolution \cite{Livnat2013} is not machine-like in any of the above senses. First, the writing itself evolves and its evolution is fundamental to its operation---its operation is not repetitive \cite{Livnat2013}. Second, the workings of evolution are not devoid of internal conflict but rather based on it, as will be discussed shortly. Third, the production of novelty is at the essence of evolution (Notice, however, that while evolution is not machine-like, its products are machine-like: evolution is a process of automatization). 

What is the nature of the writing of mutations then? As discussed in \cite{Livnat2013}, the mutation writing phenotype has the same meta-structure as that of the performing phenotype in the following sense. Take locomotion for example: we share with bears the fact that we have four limbs; but unlike bears we are habitual bipeds; and each of us may have a specific leg length and muscular details slightly different from those of others. In other words, a trait consists of widely shared and generally defined characteristics along with more specific and more narrowly shared characteristics, up to and including individual differences. According to interaction-based evolution, the mutation writing phenotype is the same in this regard \cite{Livnat2013}. It consists of generally defined and widely shared characteristics (for example, the long-term trend of the movement of genes out of the X chromosome in \textit{Drosophila} \cite{VibranovskiEtal2009}), along with more specific and narrowly shared characteristics up to and including individual differences in mutational tendencies \cite{Livnat2013}. This meta-structure explains the observations on genetic relatedness in mutational tendencies discussed in \cite{Livnat2013}. Furthermore, it implies that the nature of mutational mechanisms can be conceptualized by analogy to ecological interactions: The writing of mutations happens not according to a fixed ``rule'' but by the ever evolving ``rules of the jungle.'' This ``jungle'' is a complex one consisting of DNA and other biomolecules. The actors in it---the genetic influences on mutation---meet in an individual due to sexual reproduction, and genetic changes happen in accordance with the usual tendencies of the actors 
as well as in accordance with their individual characteristics and the particular combination they appear in in the given situation. All the while, the actors themselves slowly evolve in the long-term. Thus, when we talk about the workings of mutation, we are not talking about a harmonious, repetitive operation of a single mechanism. Instead, we are talking about the workings of an ``ecology,'' except that the outcome is not remembered in terms of energy transfers such as food-web interactions but in terms of symbolic changes in genomes. It is an ecology of information.  

According to this picture, the biological world has two facets to it, two ``forces'': one that is due to biological interactions that make their mark through differential survival and reproduction; and one that is due to biological interactions that make their mark through the writing of genetic changes \cite{Livnat2013}. These latter biological interactions are not limited to molecular mechanisms operating inside the germ cells, but involve also everything else that affects the writing of mutations, such as mechanisms of mate choice and of the sexual shuffling of the genes \cite{Livnat2013}.  

These two forces come together in the individual: the selection of individuals determines which alleles will be passed on, and the writing of alleles determines which alleles will be there in the first place. Thus, selection and writing are equally influential opponents, and they both participate in changing the genetic and phenotypic nature of the organism and thus of themselves. While each of these forces has some long-term (phylogenetically shared) tendencies, each is oblivious to the present, immediate workings of the other: the intra-organismal writing of mutations that takes place at the present moment is shielded from the external workings of natural selection that takes place at the present moment, and likewise the latter is unaffected by the former,  even though the consequences of each will eventually affect the nature of the other. In that sense, evolution arises from a conflict, or a process of negotiation, between these two fundamental forces, and what happens in the long-term must be more or less congruent with both.

\subsection{A balance of continually evolving mutational forces is responsible for genetic change}
\label{balanceofforces}

I argued above that the writing of mutations is analogous to an ecology. An ecology is a system of conflicting forces, where each species presses to produce more of itself while at the same time undoing the growth of others. And indeed, when we look at molecular evolutionary changes, we often see long-term processes that are the result of a balance of forces in the short-term. 

As an example, consider tandem gene duplication. Due to the nature of the mutational mechanisms of tandem duplication and deletion, a gene that is duplicated at tandem experiences an increased chance not only of being further duplicated but also of losing a copy \cite{GraurLi2000}. At the same time, mutations that arise in the copies in the course of evolution push toward evolutionary divergence of the copies and thus toward the cessation of duplication/deletion (because homology is required for tandem duplication/deletion), while gene conversion events push to make the copies the same again, a situation where copies are more likely to disappear. Evolution here is a reversible process where the long-term outcome depends on a balance of forces. Note also that, in this case, gene conversion may be seen as simplification, and diversification as complexification, and that the opposing tendencies to duplicate and specialize on the one hand and to equalize and collapse on the other may be part of maintaining a balance between over-specialization (or ``over-fitting'') and over-simplification, showing the importance of the ``ecology of information'' for evolution.

As another example, consider the evolution of CpG content, which plays a role in gene regulation and therefore development \cite{SuzukiBird2008,DeatonBird2011}. The cytosine in CpG dinucleotides mutates into thymine at a high rate after it is methylated \cite{HodgkinsonEyre-Walker2011}, causing CpG-poor islands to lose their CpGs \cite{MendelsonCohenEtal2011}. Importantly, this cytosine is methylated by complex enzymatic processes \cite{KloseBird2006}, which means that the locations of these mutations are determined biologically, not by accident \cite{Livnat2013}. At the same time, another mutational force---that of biased gene conversion (BGC)---adds cytosine to some CpG-poor islands \cite{GaltierEtal2001,DuretGaltier2009}, and it has been shown that the balance of such forces determines the direction of the evolution of CpG content \cite{MendelsonCohenEtal2011}. We have here a balance of mutational forces that in the long term affects functional, adaptive structure \cite{SuzukiBird2008,DeatonBird2011}, which fits with interaction-based evolution  \cite{Livnat2013} but is hard to explain from the traditional view of evolution\footnote{Indeed, as Duret and Galtier argued, CpGs work en masse, and the impact of any one particular CpG mutation is insignificant and could not be explained by traditional natural selection \cite{DuretGaltier2009}.}. Indeed, CpG mutations are not a rare anomaly, but have been estimated to account for nearly 25\% of all point mutations in humans\footnote{This is the case even though CpG dinucleotides account for only about 1\% of the human genome. Furthermore, they are often accompanied by mutations in nearby bases \cite{QuEtal2012,WalserEtal2008}, further compounding the involvement of mutational mechanisms in their origination.} \cite{FryxellMoon2005}. 

As yet another example, based on an analysis of short open reading frames in yeast, Carvunins et al.
\hspace{-4mm} 
have suggested that the evolution of new genes is gradual and reversible \cite{CarvunisEtal2012}: that a new gene does not arise suddenly as a complete whole, but gradually through forms more and more resembling a complete gene; and that at each point in time, the gene can make a step toward or away from completion. I argue that this process too is driven by a balance of forces. In mammals, for example, it would involve the evolution of CpG content. 

Finally, consider the proliferation vs. 
\hspace{-4mm} 
silencing and removal of transposable elements (TEs). There has been a well-known divide between those who think that TEs are serviceable to the organism (e.g., \cite{LynchEtal2011,Fedoroff2012,McClintock1965,BrittenDavidson1969,Shapiro2011})
and those who see them as ``selfish-elements'' \cite{Dawkins1976,OrgelCrick1980,DoolittleSapienza1980}\footnote{along with some attempts to reconcile the opposing views \cite{Doolittle1987,Doolittle2013,Livnat2013}.}: on the one hand it is now clear that TEs play an immense role in adaptive evolution 
\cite{BourqueEtal2008,SasakiEtal2008,Fechotte2008,LynchEtal2011}, and on the other hand the evolutionary ``benefit'' they bring resides in a timescale too long to allow them to fit comfortably in the traditional conceptualization of evolution except as selfish elements. However, the writing of mutations is an ecology; it is not a machine-like. TEs may well act as though they are propelled to replicate and insert themselves wherever they can, and yet, in the context of the rules of the evolving information ecology, they may be serving the evolution of the organism in the long-term\footnote{This is the case even if they have the appearance of ``selfish players'' that has been attributed to them, and even if they occasionally cause accidents in the short-term in the form of genetic disease \cite{Livnat2013}.}. Indeed, giving contra-pressure to TE proliferation is an extensive and phylogenetically deep system of regulation, involving methylation and TE removal, active in the germline \cite{ThomsonLin2009}. I argue that this extensive system is part of the ecology of mutation writing and, like Fedoroff \cite{Fedoroff2012}, I argue that it does not merely act as an ``immune defense.'' 

The four examples above clarify the view of genetic change as an ecology of information. It is a view of conflicting forces pushing against each other, including long-term reactions that may be locally reversible. This process computes in the long-term and involves the evolution of the network as a whole: the network gradually changes as it finds where it can give way under this complex set of forces. Thus, through mutational writing, the network processes a large amount of information under natural selection. 

\subsection{Evolutionary mutational mechanisms---a field open for future study} 
\label{mutationalmechanisms}

Interaction-based evolution opens up the search for non-Lamarckian yet useful mutational mechanisms. Earlier I proposed a gene-fusion mechanism that may play a role in evolution reminiscent of the role that Hebbian learning plays in neural networks (section \ref{trimcypfusion}; see also section \ref{evolutionaslearning}), according to which copies of genes that are used together are more likely to be fused together. This type of mechanism would not cause accidental changes but rather would produce evolutionarily useful genetic variation, without violating the principle that mutation does not respond to the immediate environment\footnote{Interestingly, while the mutational fusion mechanism hypothesized earlier is based on putting together empirical observations (namely the nature of transcription, chromatin states, reverse transcription and gene fusion), its general nature was predicted based on theoretical considerations of interaction-based evolution \cite{Livnat2013}: in the latter, the conceptual connection between the problem of sexual recombination and mutational mechanisms required that genes play a dual role: one of performance under natural selection, and another of influencing mutation in the germline  \cite{Livnat2013}. Thus, the coupling of germline mutation and somatic performance through transcription as in the mechanism hypothesized earlier allows for a convergence of ideas and empirically based considerations.}.

Another example of a type of mechanism that would make sense in light of the current theory is as follows. Since information about the pattern and extent of expression of a gene is present in the DNA and accessible in the germ cells in principle, a gene that is highly expressed and is therefore extensively used may be more likely to be duplicated by transcription-coupled mutational mechanisms. As in the case of the gene-fusion mechanism mentioned above, transcriptional promiscuity in the germ cells may be involved in such mechanisms (see section \ref{trimcypfusion}). When operating in unicellular organisms, such mechanisms could explain, among else, cases of rapid adaptive evolution in response to environmental pressures such as extreme temperatures, extreme salinity, or toxins, where a gene whose product is in demand is duplicated/amplified (see \cite{Kondrashovf} for review) and thus, at first glance, may seem to imply Lamarckism. However, mechanisms of this sort, coupling gene expression level to gene duplication, may serve evolution in general in a \textit{non-Lamarckian} fashion. Namely, increasing the probability of germline duplication of a gene that is extensively used in the soma may be useful because such a gene may have a greater potential to beneficially specialize evolutionarily into different functions. Thus, while in unicellulars, environmental pressures may directly cause the overexpression of a gene and thus its propensity to be duplicated through mutational mechanisms, in general it is evolution itself that would lead to the situation where a gene is highly expressed and to the application of these mutational mechanisms in a non-Lamarckian, yet useful, fashion. The fusion of \textit{CypA} and \textit{Trim5} serves as an example here as well, since this fusion involved duplication of \textit{CypA} through retrotranscription, and extensive transcription of \textit{CypA} in the germline may have facilitated this evolutionary event \cite{KaessmannEtal2009}. 

Also of interest in this regard are cases such as the evolution of insecticide resistance in the mosquito \textit{Culex pipiens} due to the amplification of the genes coding for two non-specific esterases as well as for the acetylcholinesterase that is the main target of the applied insecticides, which is active in the central nervous system \cite{LenormandEtal1998,LabbeEtal2007}. These duplications may have originated not by accident but by a gradual albeit rapid process of evolution involving natural selection and non-accidental mutation, which has created the genetic conditions under which the duplication mechanisms are more likely to be activated\footnote{Such a process would enable selection on multiple or many loci to be funneled by mutational mechanisms to influence the probability of duplication of a particular gene or genes.}. 

The current theory draws our attention to the fact that mechanisms such as the above 
can exist and puts them front and center. The examples above demonstrate that the evolving organism can receive feedback on what genetic changes would be useful to attempt---for example, what genes may be beneficially chunked together. Furthermore, this feedback comes not from the immediate environment, but from the population's past successes---the information is in the genome; there is no Lamarckism here. By accepting that mutation is not random, we can see that many findings regarding genetic activity that have been thought of before as separate phenomena may actually be working together toward a larger goal---allowing evolution to be a smart process, but one that relies on defensible principles. This view opens the door to examining future research questions that would not have come to light otherwise. 

Indeed, we may expect a diversity of mutation-writing mechanisms in nature, and the above are merely two examples of these. While some of these mechanisms  may be well known phenomena that have not yet been placed within a theoretical framework---for example the fact that recombinational mechanisms interact with DNA sequences in such manner that enables whole gene duplication and deletion---many others may remain to be discovered. 

\subsection{The intimate relationship between useful change and error repair} 

It is so often said that mutation is a replication error that one might think that this is a well-known scientific fact. However, the only fact that has actually been established is the basic observation itself---that while some genes are duplicated, others undergo genetic change. To say that these changes represent nothing more than ``replication errors'' is to provide merely one interpretation to this fact, and it may be a prejudiced one. This interpretation has led among else to the terms ``error-repair mechanisms'' and ``error-prone'' repair mechanisms, which, according to the theory presented here, may end up detracting from our understanding of evolution. 

Error is often a deviation from a pattern. By noticing a deviation from a pattern we can find and fix a typographical error: a word with a typo slightly differs from many instances of the word seen before, which are all identical to each other; and it can be fixed by making it identical to those many other instances. Then again, by noticing a deviation from a pattern, we can also avoid picking a rotten apple in the store, even if we have never seen rot or an apple before. Taking one step further, by noticing a deviation from a pattern we can also spot an error of thought. Take for example \textit{Scala Naturae}, according to which all organisms fall into a linear order from the simplest to the most advanced. From that perspective, the fact that many organisms are hard to classify as more or less advanced in relation to each other is a deviation from a pattern. By replacing \textit{Scala Naturae} with Darwin's concept of common descent, this difficulty of classification becomes not a deviation from a pattern but a part of a larger pattern involving other facts. 
Thus, both errors of typing and errors of thought can be corrected by pattern completion at different levels. At the same time, pattern completion is a form of simplification---the fewer exceptions we need to have, the smaller the amount of information needed to describe the entire system and its parts, as information theory makes clear \cite{LiVitanyi1997}. Thus, if pattern-completion operations can be implemented by mutation, we may see the same genetic mechanisms operating both in ``typographical corrections'' and in the kind of mutational writing that leads to progressive evolution. As an example, repeated events of gene conversion have the potential to correct ``typos,'' but they also have the potential to implement simplification pressure opposing the complexification pressure of diversifying in the case of duplication and deletion discussed in section \ref{balanceofforces}. 

The insufficiency of our current jargon is made particularly clear by the phrase ``error-prone repair.'' Suppose that, in some cases where so-called ``error-prone repair'' is activated, the biological system is actually pushing for a change rather than a restoration of the genetic state, and that this change is a part of a pattern-completion or other progressive evolution of the network as a whole. Then what we have heretofore thought of as ``error-prone'' is actually an attempt at ``error-correction,'' where the ``error'' is of a different, deeper kind than we use to think about. 

\subsection{Evolution as learning}
\label{evolutionaslearning}

From Paley to Dawkins \cite{Dawkins1976}, there is universal agreement that adaptations are incredibly impressive and complex pieces of ``natural technology.'' While Paley used this observation to make the non-scientific point that, much like a watch has an intelligent watchmaker, life was created by a supernatural intelligence \cite{Paley1802}, Dawkins argued that the process responsible for life is a very simple process 
of random mutation and natural selection that is fully understood at its essence. In the preface to \textit{The Blind Watchmaker} \cite{Dawkins1986},  
he wrote that ``[t]his book is written in the conviction that our own existence once presented the greatest of all mysteries, but that it is a mystery no longer... Darwin and Wallace solved it, though we shall continue to add footnotes to their solution for a while yet.'' Let us revisit the question, but from a strictly scientific perspective, and without assuming that all that is important in principle was revealed at Darwin's time: Could the process generating life forms and the process generating artificial technology 
be similar in some respects? 

Interestingly, according to Papaj \cite{Papaj1993}, it is a curious historical fact that the earliest ideas on evolution, i.e., Lamarckism, revolved around observations on \textit{automatism in behavior:} observations showing \textit{that instinct is similar to well-learned behavior}---an evolved phenomenon is similar to a learned phenomenon---in that both can be carried out automatically and independently of external influences, and both are stereotyped, or robotic, repeating with a high degree of uniformity. These observations fostered the idea that what is repeated many times over the generations gradually impresses itself upon the hereditary makeup of the organism, which then led to the additional but erroneous idea of Lamarckian transmissionism. Until now, Lamarckism has been the only alternative to natural selection at the most basic level of analysis. And even though it has been rightfully rejected as a general-level explanation for evolution, the observations it was supposed to explain are still here (and have been discussed in section \ref{innateness}). That is, the controversy was never over the \textit{observations} but rather over the \textit{mechanism} of evolution. The current paper provides a new interpretation of these original observations and suggests that there \textit{is} a connection between evolution and learning: network-level evolution and automatization are key to both. This connection is free of Lamarckian transmissionism and requires a process based on non-accidental mutation and natural selection. 

Not only in evolution, but also in the study of brain and behavior, the notion of random generation and filtering was once used. For instance, Skinner had suggested a mechanism of random generation of ideas and filtering for how learning by the brain works \cite{FodorPiattelli2011}. However, more recently, Fodor and Piattelli-Palmarini argued that such a mechanism applies neither to evolution nor to the brain \cite{FodorPiattelli2011}\footnote{Likewise, Lakatosh asked whether there is nothing more to intelligence than randomization as generating ideas and selection sifting among them \cite{Lakatosh1976}.}. In this paper and in \cite{Livnat2013}, I argued that the mechanism of evolution is not that of random generation and filtering, and that the causes of mutation are critical for our understanding of evolution. This may also inform our understanding of learning. 

In this context of connecting evolution and learning, Valiant's \cite{Valiant2009,Valiant2013} recent work attempting to connect evolution and \textit{machine} learning (see also \cite{Feldman2008,Feldman2009a,Feldman2009b,Kanade2011,AngelinoKanade2014} in the same line, and \cite{ChastainEtal2014}) signifies a methodological turning point: unlike classical population genetics, it provides rigorous mathematical techniques that capture analytically a complex phenotypic structure and allow us to quantify and study the evolution of complexity \cite{Valiant2009,Valiant2013}. Thus, with respect to theoretical methodology, it is a grand vision and, in principle, it allows mutation to be non-accidental\footnote{It allows mutation to be an outcome of any implementable randomized algorithm---an algorithm that is allowed to use random bits \textit{among else}, which is different from mutations that are nothing but random changes anywhere in the genome}. 
However, while Valiant's framework allows for, but has not yet substantially pursued, non-accidental mutation, interaction-based evolution argues that mutation \textit{is} non-accidental and that this is crucial for evolution. And while Valiant's work may be an inspiring step in the right direction, according to the present paper, there are elements that are not yet included in it that are essential for biological evolution based on non-accidental mutation. These include cooption; the idea that simplification under performance pressure produces elements that have the inherent capacity to become useful in new contexts, which leads to cooption; the idea that learning through evolutionary change is a learning from the environment by emancipation and acceleration (see section \ref{howevolutionlearns}), i.e., by the evolution of automatization and innateness; and the concept of the absorption of meaning from context under gradual network-level change. Indeed, the importance of cooption in evolution cannot be overestimated, and has been demonstrated here at both the molecular and phenotypic levels (sections \ref{contextual} and \ref{innateness}). Furthermore, cooption is analogous to an analogy or metaphor, which are crucial in the evolution of language as well as in human intelligence. It may be of much interest to explore these missing elements from a computational perspective. 

Since its inception \cite{Livnat2013}, interaction-based evolution has been deeply connected to the computational worldview \cite{Papadimitriou2007,Karp2011}, because it proposed that mutation is an event of information flow and computation: the inputs into a mutational event are the alleles at the loci affecting the mutation through genetic interaction, and the output is the mutation itself (by ``mutation'' I mean not only a change in the DNA but any heritable change, such as an epigenetic change) \cite{Livnat2013}. Furthermore, the fact that the output of a mutational event at one generation, namely the mutation itself, can serve as an input into mutational events at later generations means that the mutation-writing phenotype creates a network of information flow through the generations, from many genes into any one gene and from any one gene to many (see Figure \ref{flowovergenerations}) \cite{Livnat2013}. Other examples of networks of information flow and computation include the brain, and what computer scientists call a circuit \cite{Papadimitriou1993,Wegener1987} (one instance of which is an artificial neural network \cite{Hopfield1982}). Thus, according to interaction-based evolution, genetic evolution can be seen as the result of the workings of a network, itself evolving over time. 

Interestingly, in artificial neural networks, local computational elements are used such as Hebbian learning (e.g., \cite{Hopfield1982}). In the latter, 
when one neuron persistently participates in causing another to fire, the strength of the synapse between them is increased \cite{Hebb1949}. Hebbian learning is an example of a local simplification operation that, in the context of the gradual change of a complex network, is useful. Now, elements of this sort can play a role in the network of information flow generated by sex and non-accidental mutation proposed by interaction-based evolution  \cite{Livnat2013}; indeed, the mutational fusion mechanism in section \ref{trimcypfusion} is one such case. Thus, we see in multiple ways that, according to interaction-based evolution, evolution and ``thinking processes'' have more to do with each other than previously thought, even though no Lamarckism and no ``foresight'' or ``adaptive mutation'' as traditionally defined are involved. Thus, the study of evolution could inform the study of learning and vice-versa.  

Recently, a connection between evolution and learning was drawn by Watson and Szathm{\'a}ry \cite{WatsonSzathmary2015} and Watson et al. \cite{WatsonEtal2016}. While this connection shares with the theory of interaction-based evolution as proposed in \cite{Livnat2013} and here the idea that evolution is network-based, and that the change of connections between the nodes of the network is key, there are also some fundamental differences between the two. Watson and Szathm{\'a}ry \cite{WatsonSzathmary2015} and Watson et al. \cite{WatsonEtal2016} did not argue for non-accidental mutation, and all that follows from it. 

For example, it follows from non-accidental mutation that Hebbian-learning--like mechanisms can be implemented directly by the mutational mechanisms themselves (as opposed to needing to arise from random mutation and natural selection), as discussed in section \ref{trimcypfusion}. There, I argued that genes that are used together are fused together\footnote{In contrast, Watson et al. argue that ``genes that are selected together are wired together.'' There is a fundamental difference between the two statements, because Watson et al. imply a process of random mutation and natural selection. Namely, they base their statement on a pioneering theoretical model \cite{PavlicevEtal2011}, but one that is constructed within the random-mutation view.}. More generally, I argued that simplification can be implemented by mutational mechanisms. In fact, the very concept of non-accidental mutation itself represents a vast network, as discussed here and in \cite{Livnat2013}. Conceptualizing mutation not as a local accident disconnected from its genetic environment, but rather as the outcome of network-based processes, provides a far more involved network-based view of evolution than otherwise. It also greatly strengthens the connection between computer science and evolution \cite{Papadimitriou2007,Karp2011}.

In addition, borrowing from knowledge in machine learning, the above-mentioned authors mention that, among other things, imposing parsimony pressure by imposing a connection cost in models of genotype-phenotype maps can facilitate evolution in these models \cite{WatsonSzathmary2015,WatsonEtal2016,CluneEtal2013,KouvarisEtal2015}. However, they do not put simplification pressure front and center in biological evolution, as done here. The present paper established the importance of simplification in biological evolution by providing both the rationale and many empirical examples from both the molecular and phenotypic levels behind this point. On this foundation, it argued that biological evolution is driven by two forces---the pressure for performance and the pressure for simplification. A cycle in evolution begins at a fuzzy state from the emergent interactions between preexisting elements. From these interactions, simplification under performance pressure creates new elements that have the inherent capacity to come together into unexpected, useful interactions with other such elements. This leads to cooption, and to the beginning of another cycle in the process. Thus, putting simplification front and center in biological evolution also puts cooption at the heart of the evolutionary process. In addition, from this cycle we also obtain the idea that simplification leads to biological complexity (section \ref{simplification}). 

While one would need to justify biologically a substantial cost to a single genetic connection \textit{per se} if traditional selection is to simplify a genetic network based on random mutation, according to interaction-based evolution, simplification is inherent to biological evolution, and can be implemented by mutational mechanisms. Indeed, interaction-based evolution argues that mutational mechanisms, mixing of the hereditary material (which has evolved into sex), simplification and selection have all existed from the ``beginning'' of life, and that they did not evolve from an asexual world with random mutation, since that world never existed \cite{Livnat2013}. Thus, they are not elements that evolved by random mutation and natural selection based on different costs and benefits imposed by selection, but rather are primary elements that are original and inherent to the process of evolution. Thus, interaction-based evolution is different from the evolvability view present in previous biological literature. 

Indeed, interaction-based evolution provides a complete, biologically motivated, conceptual framework for evolution with non-accidental mutation at its center. By arguing that novelty arises from emergent interactions, it places the source of novelty at the system level. This in turn replaces the notion of accidental mutation as the ultimate source of heritable novelty, which in turn connects back to the center-piece of non-accidental mutation. This entire framework and all of its elements, including cooption, novelty and non-accidental mutation, as well as the idea that simplification leads to complexity, and the idea that evolutionary learning occurs through automatization and innateness, have not been discussed in previous papers on evolution and learning. 

\section{Conclusions}

How does novelty arise? Traditional evolutionary thinking relies on random mutation and natural selection. The idea is that radiation, or a copying error, or oxidative stress, ``goes zap,'' and a new mutation appears that, on rare occasions, provides a beneficial phenotypic change. All that remains for natural selection to do is to check whether this mutation on its own is good or bad---to play the role of a filter. Where does the novelty, the new genetic information, come from? Presumably, in that view, it comes from the accident itself---from out of thin air---and there is nothing more to inquire regarding the source of it. 

Interaction-based evolution proposes an alternative to this view. The mutations that are relevant for adaptive evolution under selection are due to mutational mechanisms that are continually evolving, and that do not in and of themselves invent things. Rather, novelty arises from the system level---from the macroscale---from gradual network-level evolution, as these mechanisms absorb information from selection. In brief, mutation mechanisms perform simplification operations on the genetic network, as well as gene duplication, in a heritable mode. These mechanisms work together with natural selection which acts on the organism as a complex whole, so that adaptive evolution is a process of simplification under performance pressure. A cycle in this process begins with complex high-level interactions between preexisting elements. Simplification under performance pressure takes these preexisting interactions and, gradually, in the course of evolution, creates from them new elements---new adaptations. Because these new elements are created in a process of simplification under performance pressure, they have the inherent capacity of coming together in new, useful and unexpected interactions at higher levels, thus initiating another cycle in the process. This capacity to come together in useful high-level interactions that have not been pursued in advance is the source of novelty in evolution. In short, mutations do not in and of themselves invent things, but rather are a key activity that takes part in turning the wheel of evolution. Interestingly, it is simplification that explains complexity: local simplification leads to a global increase in complexity. 

Thus, while traditional theory is based on the idea that random mutation invents---where this supposed random mutation is a remote, presumed event that cannot be seen or confirmed---the theory presented here is based on the empirically evident  fact that preexisting elements come together into new, useful high-level interactions as the source of novelty in evolution. Note that it matters not whether the novelty involved in the transitions from the genes \textit{TRIM5} and \textit{CypA} to their fusion, or from haphazard egg retrieval to backward walking, is small or great in and of itself. Rather, these transitions exemplify the steps that tie together the process of evolution, which in the long term lead from the progenote to humans. 
 
We have a tendency to look for ``foundations'' from which everything else can be derived. In particular, it is convenient to assume that the causes of mutation are random, because it puts an end to all of our questions. The philosophical move that is required from the perspective of interaction-based evolution is to let go of the notion that random mutation and novelty from a point are at the bottom of things---that they provide a stable ground upward from which a conceptual edifice can be built; and to accept instead that the action is at the network level: that both the meaning \textit{and origin} of genetic and phenotypic elements comes from the higher levels of organization---it comes from the network---from ``above.'' This move opens up the study of evolution substantially; because while the notion of random mutation means that there is nothing of importance to be studied about the causes of mutation from an evolutionary perspective, the concept of non-accidental mutation provided by interaction-based evolution implies instead a whole world of biological mechanisms open to investigation. 

Before Darwin, people used to think that different species were each created separately in an instant. While Darwin made an immense contribution by showing that this was not the case, and that species are generated gradually, a notion of creation in an instant has been maintained in neo-Darwinism in other areas: the origin of life, the origin of mutations, and cooption. While \cite{Livnat2013} argued among else against the origin of life as an instant, this paper argues against the other two. Novelty arises not suddenly from a point but from gradual network-level evolution. Indeed, if evolution according to random mutation and natural selection is a sequence of independent points, each representing a local accidental mutation disconnected from the rest, interaction-based evolution draws the lines between these points (see \cite{Livnat2013}) while fundamentally altering their interpretation. 

\section{Acknowledgements}
I would like to thank Marc Feldman, Avraham Korol, Simon Levin, Amos Livnat, Daniel Melamed, Steve Pacala, Christos Papadimitriou, Nick Pippenger, Umesh Vazirani and Kim Weaver for invaluable conversations during the course of the study. I would like to thank the Department of Evolutionary and Environmental Biology and the Institute of Evolution at the University of Haifa for providing an intellectual environment conducive to the pursuit of this work, and for stimulating conversations. I would like to acknowledge financial support from the Miller Institute for Basic Research in Science and from NSF grant 0964033 to Christos Papadimitriou, Division of Computer Science, UC Berkeley, for support during a formative part of the work in the years 2006--2011. 

\bibliographystyle{abbrv}
\bibliography{LivnatReferences3}

\vspace{2 mm}

\newpage
\thispagestyle{empty} 

\end{document}